\documentclass{psp-rv975x65}
\usepackage{psp-rv-van}             % numbered citation/references (default)
\usepackage{subfigure}
\newcommand{\be}{\begin{equation}}
\newcommand{\ee}{\end{equation}}
\newcommand{\ep}{\epsilon}
\usepackage{color}
\usepackage{float}

\makeindex
\copyline{Book Title}{2009}{978-981-nnnn-nn-n}
\begin{document}

\chapter[]{Wave propagation in one-dimension: \\ Methods and applications to complex and fractal structures \label{ch1}}

\author{E.~Akkermans$^{(1)}$, G.~V.~Dunne$^{(2)}$, and  E.~Levy$^{(1)}$ }

\address{$~^{(1)}$ Department of Physics, Technion, Israel Institute of Technology, 32000, Haifa, Israel\\
$~^{(2)}$ Department of Physics, University of Connecticut, Storrs CT 06269, USA}

\begin{abstract}
This chapter is a pedagogical review of methods and results for studying wave propagation in one-dimensional complex structures. We describe and compare the tight-binding, scattering matrix, transfer matrix and Riccati formalisms. We present  examples for transport through finite-sized layered dielectric systems with periodic, quasi-periodic, fractal, disordered, and random structure, illustrating how can spatial structure affect the spectrum of modes as well as the local mode intensity.
\end{abstract}

\body

\section{Introduction}

Both classical and quantum waves are highly sensitive probes of the details of the physical and geometric structure of the medium in which they propagate. Periodic structures produce spectra with transmission bands, leading to the remarkable physics of photonic band gaps \cite{pbg}. Other kinds of structure, quasi-periodic \cite{marcia,damanik}, fractal, disordered or random \cite{erdos,gredeskul,luck,ambook}, lead to other distinct spectral features, some of which suggest certain advantages for physical and engineering applications. Important questions include the frequency distribution of modes, as well as their spatial localization. Recent experimental advances suggest that various types of complex media can now be fabricated with high precision. This review is designed to summarize some basic theoretical tools for designing layered structures in order to produce desired properties of the mode spectra and the local mode densities.

We describe the density of states, the counting function (the integrated density of states), the transmission probability, and the local field amplitude and intensity, each of which has a direct physical meaning. We compare and contrast different computational methods: the tight-binding, scattering matrix, transfer matrix, and Riccati methods, and we present a variety of results to illustrate the effect of the layered structure type on the aforementioned physical quantities.

\section{Wave equations}

The general  aspects of coherent propagation are common to  a wide variety of waves
which propagate in scattering media. This notwithstanding, each
type of wave exhibits its own characteristic behavior.  In this
section, we present several examples of wave equations, and we
study two important classes in greater detail~: the Helmholtz equation,
which describes scalar wave propagation suited for electromagnetic TE or TM modes
propagating in a dielectric, and the Schr\"odinger
equation associated with a non interacting electron gas (weakly
disordered metals or semiconductors).

\subsection{Helmholtz equation}

The case of electromagnetic  waves is special, for several
reasons. It is probably one of the earliest examples
where changes in wave phase coherence due to passage through a random
medium was examined.  In the beginning of the twentieth century, very precise
studies were carried out on electromagnetic wave propagation
through diffusive media, specifically the atmosphere. From a
conceptual viewpoint, this problem stimulated the community
working in the theory of probability, who regarded it as a new
field for the application of methods developed for the study of
Brownian motion \cite{Frisch68}.  For the atmosphere, the
description in terms of a static disordered medium is not
appropriate.  For many other cases, however, the description in
terms of static disorder described by a time-independent potential
works well, and it is this case that we consider in the present
section.

For a TE mode propagating in the $x$ direction along a medium with spatially varying dielectric function $\epsilon(x)$, Maxwell's equations for the electromagnetic field amplitudes, $E_z = \psi (x) e^{-i \omega t}$ and $H_y = \chi (x) e^{-i \omega t}$, become two coupled equations \cite{Budden}:
\begin{eqnarray}
{d \psi \over dx} =i k_0\, \chi \quad, \quad
{d \chi \over dx} = i k_0\, {\ep (x) \over {\overline \ep}} \psi
\end{eqnarray}
where $k_0={\overline n} \omega/c$, namely for an average dielectric of refractive index ${\overline n} = \sqrt{ {\overline \ep} / \ep_0}$ and average dielectric constant $\overline \ep$ (See for instance \cite{ambook}, Chap.2).
%where $\ep(x)=\ep_0$.
This reduces to a scalar Helmholtz equation for $\psi(x)$:
\be
- {d^2 \psi \over dx^2} - k_0^2\, v(x) \psi(x) = k_0^2\, \psi (x)
\label{helm}
\ee
where $v(x) \equiv {\ep (x) \over \overline \ep} -1 \equiv n^2 (x) -1$, with  refractive index $n(x)$.
%,  and the wave vector $k_0 = \sqrt{ {\overline \ep}} \omega / c$. The wave equation (\ref{helm}) thus rewrites
%\be
%{d^2 \psi \over dx^2} + {\omega^2 \over c^2 } n^2 (x) \psi = 0
%\label{eq3}
%\ee
%The previous form results also from the evolution of the transverse components $E_z$ and $H_y$ of the electromagnetic field given by $E_z = \psi (x) e^{-i \omega t}$ and $H_y = \chi (x) e^{-i \omega t}$ which obey the two coupled first order equations:
%\begin{eqnarray}
%{d \psi \over dx} &=& i k_0 \chi \nonumber \\
%{d \chi \over dx} &=& i k_0 \ep (x) \psi
%\end{eqnarray}
%Eliminating either $\psi$ or $\chi$ produces the Helmholtz equation (\ref{eq3}).

\subsection {Schr\"odinger equation}

The Schr\"odinger wave equation for a particle of mass $m$  in a one dimensional potential $V(x)$ also has a Helmholtz-like form:
\be
- {\hbar^2 \over 2m } {d^2 \psi \over dx^2} + V(x) \, \psi = E \, \psi (x)
\label{schrod}
\ee
This can be formally written as $H \, \psi = k^2 \, \psi$.

The two examples we have discussed are not the only ones to
exhibit effects related to coherent propagation in complex media.  In fact, these effects
are common to all wave phenomena (quantum, optical, hydrodynamic,
etc), independent of dispersion relation and space dimension,
provided that there are no nonlinear effects.  Indeed, these may
hide complexity effects.  Moreover, nonlinear equations often have
special solutions (solitons, vortices,...) whose stability is
ensured by a topological constraint which is very difficult to
destabilize by means of a disordered potential. The competitive
role of disorder and nonlinearity is important but is still relatively poorly understood
\cite{nonlineaire}.

%TIGHT BINDING- DISCRETE VERSION

\section{Tight binding formalism}
\label{sec:tight}

For certain applications, a discrete version of the Helmholtz  and Schr\"odinger equations may be more appropriate or more tractable numerically. The tight-binding version of the Schr\"odinger equation is widely used, for example in condensed matter physics \cite{luck,tight}. We write (\ref{schrod}) in the general form:
\be
{ 1\over 2} {d^2 \psi \over dx^2} - v(x) \psi = - k^2 \psi
\label{tb1}
\ee
Discretize space into a succession of identical layers of thickness $\Delta$:
\be
{1 \over 2 \Delta^2 }  \left[ \psi_{n+1} ( k) -  2 \psi_n( k) + \psi_{n-1}( k) \right]
+ \left[ k^2 - v_n \right] \psi_n ( k) =0
\label{tb2}
\ee
In the limit $\Delta \rightarrow 0$, the previous equation can be rewritten as:
\be
\frac{1}{2 \Delta^2 } \psi_{n+1}( k) + \frac{1}{2 \Delta^2}  \psi_{n-1}( k)
=\left( 1 -  k^2 \Delta^2 \right) e^{ \Delta^2 v_n} \psi_n( k)
\label{tb3}
\ee
Defining the dimensionless quantities: $e = 1 - k^2 \Delta^2$, $\phi_n (e) = e^{\Delta^2 v_n / 2} \psi_n$ and $t_{n,n+1} = e^{ - (\Delta^2 /2) ( v_n+ v_{n+1})}$, the wave equation takes the canonical discrete form:
\be
t_{n,n+1} \,\phi_{n+1} (e) + t_{n,n-1}\, \phi_{n-1} (e) = 2 \,e\, \phi_n
\label{tb4}
\ee
often referred to  as {\it non-diagonal} tight-binding description. The equivalent {\it diagonal} tight-binding form is:
\be
\phi_{n+1} (e) +  \phi_{n-1} (e) + v_n \, \phi_n = 2\, e\, \phi_n
\label{tb5}
\ee
These two equivalent forms define three-term recursion relations \cite{3terms}, whose solutions are families of polynomials.  They show up in various problems such as the master equation, random walks, and laplacian problems. An important issue is to determine the corresponding family of properly normalized orthogonal polynomials and their spectral distribution. For example, the solutions for a periodic potential are in terms of Chebychev polynomials, as discussed below in Section \ref{secperiodic}.

For a system of length (number of sites) $N$, the tight-binding equation can be written in a matrix form $H_N \Phi = e\, \Phi$. The system has thus $N$ eigenenergies,  we shall denote by $e_1, \dots, e_N$. The {\it counting function}, $\mathcal{N}(e)$, or {\it integrated density of states}, is defined as the fraction of eigenenergies that are smaller than a given energy $e$:
\be
\mathcal{N}(e) = {1 \over N} \sum_{i=1}^N \theta (e - e_i )
\label{counting}\ee
where $\theta (x)$ is the Heaviside step function, equal to $1$ (resp. 0) for $x \geq 0$ (resp. $x <0$). For large enough $N$, the counting function is independent of the choice of boundary conditions. The counting function is usually a well defined and continuous function of energy. Its derivative, $\rho (e)=d \mathcal{N}/de$, if it exists, is called the {\it density of states} (or {\it density of modes}). It is a sum of Dirac delta functions,
\be
\rho (e) = { 1 \over N} \sum_{i=1}^N \delta (e - e_i ) \, .
\label{dos}\ee
As we shall see, for fractal systems, the counting measure $d \mathcal{N}(e)$ exhibits a singular behavior related to divergences of the density of states.
\medskip

{\bf Remark :}
There is a connection between the counting function and the zeroes of the wave function in one-dimensional systems. Consider for simplicity  the form (\ref{tb5}) of the tight-binding wave equation for $N$ sites. Assuming zero boundary conditions (Dirichlet), $\phi_0 = \phi_{N+1} =0$, and $\phi_1 =1$ as a normalization, we have, $\phi_2 = e - v_1$, $\phi_3 = (e-v_1)(e-v_2) -1$, and so on, so that,
\be
\phi_{N+1} = \prod_{i=1}^N (e-e_i ) \, ,
\ee
namely the zeroes of the wave function are related to the location of eigen-energies.
\medskip

The tight binding description allows to make interesting connections to other descriptions such as the transfer matrix approach (see sections \ref{sec:tmatrix} and \ref{secperiodic}). A simple illustration for the connection is the simple case of the laplacian evolution in the absence of scattering potential. It corresponds to (\ref{tb1}) for $v(x) \equiv 0$. For the tight binding problem it corresponds to taking $t_{n,m} \equiv 1$ for all sites, namely to solve the three term recursion \cite{3terms}:
\be
\phi_{n+1} (e)  = 2 e \, \phi_n -  \phi_{n-1} (e)  \, .
\label{bloch1}
\ee
The solution of this specific recursion is well known and given by the Chebyshev polynomials of the first kind. For $-1 \leq e \leq 1$, the solutions are $\phi_n (e) = \phi_n (\cos \theta ) = T_n (\cos \theta ) = \cos n \theta$, where $e = \cos \theta$, and $\theta \in [ - \pi , \pi ]$ is the so called Bloch angle. Their spectral distribution is characterized by the density of states:
\be
\rho (e) = { 1 \over 2 \pi} {d \theta \over de} = { 1 \over 2 \pi } {1 \over \sqrt{1 - e^2}}
\label{dostb}
\ee
which corresponds to a well known results both in tight-binding and for the invariant spectral measure of Chebyshev polynomials \cite{3terms}.

\subsection{Tight-binding approach to the Helmholtz equation}
\label{tbhesection}

The tight binding version of the scalar wave (Helmholtz) equation is obtained along the same lines \cite{tight}. To derive it, consider instead of the continuous variation of the refractive index $n(x)$, a succession of $N$ dielectric layers extending from $0 \leq x \leq x_1$, $x_1 \leq x \leq x_2$, $\cdots$, $x_{N-1} \leq x \leq x_N$. For each layer of thickness $l_i$ there is a constant refractive index $n_i$. We can then rewrite (\ref{tb3}) as:
\be
\psi_{i+1} = \lambda_i \psi_i - \mu_i \psi_{i-1}
\label{tbh}
\ee
where
\begin{eqnarray}
\lambda_i &=& {n_i \sin \beta_{i+1} \over n_{i+1} \sin \beta_i} \cos \beta_i + \cos \beta_{i +1} \nonumber \\
\mu_i &=& {n_i \sin \beta_{i+1} \over n_{i+1} \sin \beta_i}
\label{tbhcoeff}
\end{eqnarray}
and $\beta_i = k_0 n_i l_i$.

\section{Scattering matrix formalism}
\label{sec:smatrix}

\begin{figure}[H]
%\centerline{\includegraphics[scale=0.4]{smatrix.eps}}
\centerline{\includegraphics[scale=0.4]{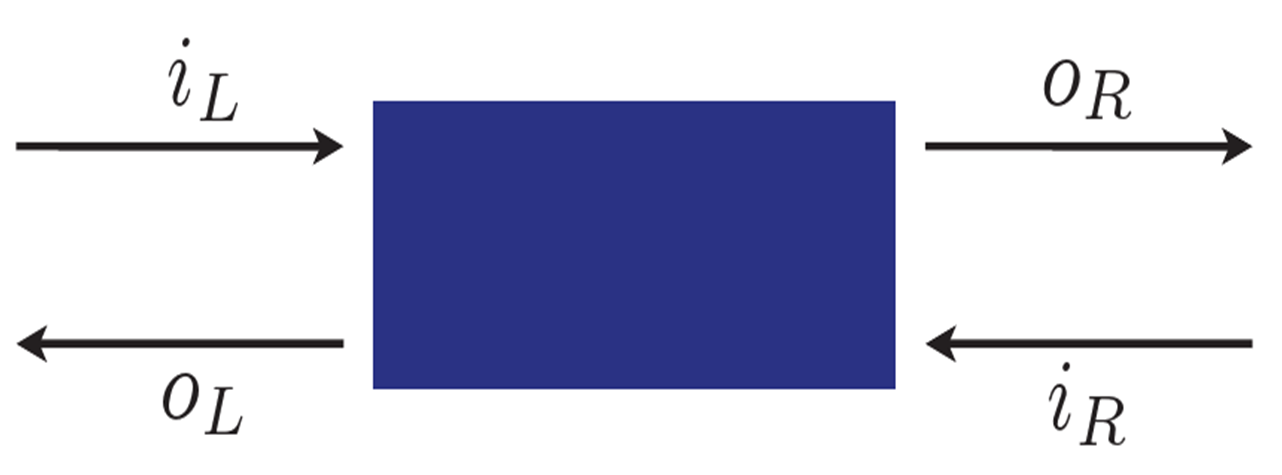}}
%\centerline{ \epsfxsize 7cm \epsffile{9.eps}
 \caption {\em Schematic description of the scattering matrix setup, showing the incoming ($i_R,i_L$) and outgoing ($o_R,o_L$) amplitudes on the left and right of the scattering region.}
 \label{fig1}
\end{figure}

Propagation through a one-dimensional structure can be described by a scattering matrix, or S-matrix, $S(k)$, relating incoming and outgoing amplitudes of propagating plane waves of wave vector $k = \omega / c $ (see Figure \ref{fig1}) . Excellent pedagogical discussions, particularly for one-dimensional systems, can be found in \cite{avishai,konishi,visser}.
%------------------------------
%\begin{figure}[ht]
%\includegraphics[scale=0.55]{SMMatrix.png}
%\caption{Schematic description of the scattering $(S)$ and transfer $(M)$ (see section \ref{sec:tmatrix}) matrices setups.}
%\label{figSMMatrix}
%\end{figure}
%------------------------------

%------------------------------
%------------------------------

With obvious notations, the scattering $S$ matrix is defined as:
\begin{eqnarray}
 \left(
\begin{array}{l}
o_L  \\
o_R
\end{array}
 \right)
 =  \left(
\begin{array}{ll}
r \, & t \,  \\
t \,  & r' \,
\end{array}
 \right)  \left(
\begin{array}{l}
i_L  \\
i_R
\end{array}
 \right)
 \equiv S\,
 \left(
\begin{array}{l}
i_L  \\
i_R
\end{array}
 \right) \, .
\label{smatrix0}
\end{eqnarray}
We consider the system to be invariant under time reversal, so that the matrix $S$ is symmetric. Furthermore,  it is unitary ($S^{-1}=S^\dagger$) as a consequence of conservation of probability (for the Schr\"odinger equation), or of the intensity of the field (for the Helmholtz equation). This leads to the set of relations:
\begin{eqnarray}
|r|^2 + |t|^2 &=& 1 \\
|r'|^2 + |t|^2 &=& 1 \\
t r^* + r' t^* &=& 0 \, .
\label{unitary}
\end{eqnarray}
These equations imply that $\mbox{det}\, S = rr' - t^2 = - t / t^*$. Since $S$ is unitary, it can be diagonalized by a unitary transformation into the diagonal form:
\be
\left(
\begin{array}{ll}
e^{i \phi_1}& 0 \\
0 & e^{i \phi_2}
\end{array}
 \right) \, .
 \label{sdiagonal}
 \ee
 Defining the {\it total phase shift}, $\delta (k) \equiv (\phi_1 (k) + \phi_2 (k))/2$, we then have:
 \be
 \mbox{det}\, S (k)= e^{2 i \delta (k)} = - {t \over t^*} \, .
 \label{dets}
 \ee
From the definition of the phase of the transmission amplitude, $t\equiv |t|e^{i \, \alpha}$, and from (\ref{dets}), we obtain the relation $\delta(k)=\alpha (k) +\pi/2$. A simple and elegant relation exists between the phase shift $\delta(k)$ and the change of density of states $\rho(k)$:
 \be
\rho (k) - \rho_0 (k) = {1 \over 2 \pi}\, \mbox{Im} \,\frac{\partial}{\partial k} \ln \mbox{det}\, S(k)
\label{theequation}
\ee
where $\rho_0(k)$ is the density of states for the free system, namely with zero potential for the Schr\"odinger equation in (\ref{schrod}), or $\ep(x)=\ep_0$ for the Helmholtz equation (\ref{helm}).

The source of scattering in either Schr\"odinger or Helmholtz equations is the potential $V(x)$ (or the varying refractive index $n(x)$). Should they vanish, the $S$-matrix reduces to the identity. Now assume that they decrease fast enough so that we can enclose the scattering system (the ``black box'' in Fig. \ref{fig1}) inside a region of size $L$, much larger than the support of the scattering potential.
% (see Fig. \ref{fig2}).
 Apply periodic boundary conditions, $\psi  (0) = \psi (L)$ and $\psi ' (0) = \psi ' (L)$, at the boundary of the large box, noting that for large enough $L$, the physics is independent of the precise boundary conditions.
 For large enough $L$,
 \begin{eqnarray}
\psi (0) =\psi (L) &\quad \Rightarrow\quad &
 i_L  + o_L = o_R e^{ikL} + i_R e^{-ikL} \nonumber \\
\psi^\prime (0) =\psi^\prime (L) &\quad \Rightarrow\quad &
ik(i_L - o_L )=ik(o_R e^{ikL} - i_R e^{-ikL}) \, .
\label{bcs}
\end{eqnarray}
These algebraic relations may be written as a spectral condition:
\begin{eqnarray}
\mbox{det} \left(  1 - e^{ikL}  \left(
\begin{array}{ll}
0& 1 \\
1 & 0
\end{array}
 \right) S(k) \right) &=& 0 \, .
% \nonumber \\
% \mbox{det} \left(  1 -  \left(
%\begin{array}{ll}
%0& e^{-ikL}  \\
%e^{ikL}  & 0
%\end{array}
% \right) M(k) \right) &=& 0
\label{spectral1}
\end{eqnarray}
Solving for $S$, leads to the following relation between the total phase shift defined in (\ref{dets}) and the possible wave vectors:
\be
 k_n (L) = {\pi n \over  L} - {\delta (k_n) \over  L}
\label{trans}
\ee
Noting that $k_n ^{(0)} (L) = \pi n / 2 L$ are the eigenmodes in the absence of scattering potential, namely for $\delta (k) = 0$, we can rewrite (\ref{trans}) for two consecutive values of $n$ under the form,
\be
( k_{n+1} - k_n )  \left( L + {d \delta (k) \over dk} \right) = \pi
\ee
Defining the density of states, or density of modes (DOM) as $\rho (k) = 1 / ( k_{n+1} - k_n )$ leads to (\ref{theequation}):
\be
\rho (k) - \rho_0 (k) = { 1\over \pi} {d \delta (k) \over dk}
\ee
Recalling the definition in (\ref{counting})  of the counting function, $\mathcal{N}(k) = \int^k \rho(k') dk'$, and its relation to the DOM in (\ref{dos}), we obtain:
\be
\delta (k) = \pi \Delta N(k) = \pi \left( N(k) - N_0 (k) \right)
\label{counting18}
\ee
which relates the total phase shift of the $S$-matrix to a spectral quantity. Both (\ref{theequation}) and (\ref{counting18})  are rather remarkable (and well-known) results since they express the fact that a measurement of the scattering data from a black box allows one to retrieve its spectral information provided it is coupled to the external environment. Some further details are given in the next section.
%These results known back to Schwinger and others have been formalized in details and we present some of it in the next section.

\subsection{Derivation 2: Green's function and resolvent}
\label{sec:greens}

%The previous derivation is rather general but not very systematic. The relevant general framework valid for all types of waves relies on the use of Green's functions.
A more formal derivation, valid for all types of waves, is formulated in terms  of Green's functions
\cite{newton}. A thorough description for the cases of interest here is presented (among others) in Chapter 3 of \cite{ambook}. Here we present some salient results as well as new expressions obtained using the Gelfand-Yaglom description. For practical purpose, we consider the Schr\"odinger equation, but we keep in mind that these considerations apply equally to other wave equations. We denote by $H$ the Hamiltonian of the scattering system and by $H_0$ those of the free system, so that $V = H - H_0$ defines the scattering potential.

We define the Green's operators (resolvents) for the systems with and without the potential:
\begin{eqnarray}
G_\pm=\frac{1}{E\pm i\varepsilon -H} \quad , \quad G_\pm^{(0)}=\frac{1}{E\pm i\varepsilon - H_0}
\quad, \quad \varepsilon\to 0^+
%G_\pm^{(0)}=\frac{1}{E\pm i\epsilon - H_0}\quad , \quad G_\pm=\frac{1}{E\pm i\epsilon -H}
\label{greens}
\end{eqnarray}
The relation to density of states $\rho (E)$ defined in (\ref{dos}) follows from the identity:
\begin{eqnarray}
G_+-G_-&=&\frac{-2i\varepsilon}{(E-H)^2+\varepsilon^2}=-2\pi \, i\, \delta(E-H)
%\nonumber\\
%G_+^{(0)}-G_-^{(0)}&=&\frac{-2i\epsilon}{(E-H^{(0)})^2+\epsilon^2}=-2\pi \, i\, \delta(E-H^{(0)})
\label{dos22}
\end{eqnarray}
%Further, we can write
%\begin{eqnarray}
%G_\pm =-\frac{\partial}{\partial E}\, \ln \, G_\pm
%%\quad, \quad G_\pm^{(0)} =-\frac{\partial}{\partial E}\, \ln \, G_\pm^{(0)}
%\label{logg}
%\end{eqnarray}
Thus, we have
\begin{eqnarray}
\rho(E)&\equiv& {\rm tr}\,\delta(E-H)=- \frac{1}{2\pi i} {\rm tr}\left( G_+-G_-\right)\nonumber\\
&=& \frac{1}{2\pi i} \frac{\partial}{\partial E}\, {\rm tr}\, \ln\, \frac{G_+}{G_-}= \frac{1}{2\pi i} \frac{\partial}{\partial E}\, \ln\, \det\, \frac{G_+}{G_-}
\end{eqnarray}
%Therefore, clearly, assuming $G_+=G_-^\dagger$ [i.e. $H$ hermitean and $E$ real]
%\begin{eqnarray}
%{\rm Im}\, G_\pm&=&\mp\pi\, \delta(E-H)\nonumber\\
%{\rm Im}\,G_\pm^{(0)}&=&\mp\pi\, \delta(E-H^{(0)})
%\label{dos2}
%\end{eqnarray}
%Express G in terms of log G:
%\begin{eqnarray}
%G_\pm =-\frac{\partial}{\partial E}\, \ln \, G_\pm\quad, \quad G_\pm^{(0)} =-\frac{\partial}{\partial E}\, \ln \, G_\pm^{(0)}
%\label{logg}
%\end{eqnarray}
%Density of states in terms of log G:
%\begin{eqnarray}
%\rho(E)&\equiv& {\rm tr}\,\delta(E-H)\nonumber\\
%&=& \mp \frac{1}{\pi} {\rm tr}\, {\rm Im}\, G_\pm\nonumber\\
%&=&\pm \frac{1}{\pi} {\rm tr}\, {\rm Im}\, \frac{\partial}{\partial E}\, \ln\, G_\pm\nonumber\\
%&=& \frac{1}{2\pi} {\rm tr}\, {\rm Im}\, \frac{\partial}{\partial E}\, \ln\, \frac{G_+}{G_-}
%\end{eqnarray}
Here we have used the fundamental relation:
\begin{eqnarray}
{\rm tr}\, \ln ({\rm operator})=\ln\,\det ({\rm operator})
\label{logdet}
\end{eqnarray}
which is clear for diagonalizable matrices, but which also applies to the operators being considered here\cite{newton,simon}.
In fact, the density of states should be defined relative to the free density of states:
\begin{eqnarray}
\Delta\rho(E)=\rho(E)-\rho_0(E)&=&\frac{1}{2\pi i} \frac{\partial}{\partial E}\, \ln\, \det \left( \frac{G_+}{G_+^{(0)}}\frac{G_-^{(0)}}{G_-}\right)\nonumber\\
&=&\frac{1}{2\pi i} \frac{\partial}{\partial E}\, \ln\,\frac{\det\left(\frac{G_+}{G_+^{(0)}}\right)}{\det\left(\frac{G_-}{G_-^{(0)}}\right)}
\label{dostates}
\end{eqnarray}
For a hermitean Hamiltonian $H$,  the S-matrix  is defined as \cite{newton}
\begin{eqnarray}
S\equiv \frac{G_+}{G_+^{(0)}}\frac{G_-^{(0)}}{G_-}\quad \Rightarrow \quad S^\dagger=S^{-1}\quad ({\rm unitarity})
\label{smatrix}
\end{eqnarray}
and so the result (\ref{theequation}) follows.
%Since $S$ is unitary we can write $S=e^{i\eta}$, and then
%\begin{eqnarray}
%\rho(E)-\rho_0(E)=\frac{1}{2\pi}{\rm tr}\left( \frac{\partial\eta}{\partial E}-\frac{\partial\eta_0}{\partial E}\right)
%\end{eqnarray}
%which has the form of the Freidel sum rule expression for a single radially symmetric point-scatterer in terms of the phase shifts:
%\begin{eqnarray}
%\rho(E)-\rho_0(E)=\frac{1}{2\pi}\sum_l (2l+1)\, \frac{\partial\eta_l}{\partial E}
%\end{eqnarray}
%\\

These results are part of a vast  literature on the subject generally known as the Krein-Birman-Schwinger formulation \cite{kbs,schwinger,gold}. It is based on the generalization of the resolvent operator $G(z) = 1 / (z- H)$  to the complex $z$-plane. The determinant of the $S$-matrix defined in (\ref{smatrix}) can be related to the resolvents defined in the presence ($G$) and in the absence ($G^{(0)}$) of the scattering potential,
\be
\ln \mbox{det} S(z) = \ln \mbox{det} (z - H) - \ln \mbox{det} (z - H_0 ) \, .
\label{thouless1}
\ee Thus,
\be
\mbox{Im Tr} \left[ G (E + i 0^+ ) - G_0 (E + i 0^+ ) \right] = \pi {d \over dE} \left( {i \over 2 \pi} \ln \mbox{det} S(E) \right) \, .
\ee
Recalling that the density of states is given by $\pi \rho (E) = - \mbox{Im Tr} G(E + i 0^+ )$ leads again to (\ref{theequation}). This relation is a particular case of a more general set of relations known as the Krein-Birman-Schwinger relations which can be rewritten as,
\be
\mbox{Tr} \left[ G(z) - G_0 (z) \right] = \int_{- \infty} ^{+ \infty} {dE \over (E-z)^2} {i \over 2 \pi} \ln \mbox{det} S(E)
\label{krein1}
\ee
or more generally,
\be
\mbox{Tr} \left[ \Phi (H) - \Phi (H_0) \right] = \int_{- \infty} ^{+ \infty} dE {d \Phi \over dE} {i \over 2 \pi} \ln \mbox{det} S(E)
\label{krein2}
\ee
where $\Phi (H)$ is some regular function of the Hamiltonian $H$. When applied to $\Phi (x) = e^{-tx}$, (\ref{krein2}) gives the well known relations between the heat kernel $P(t) \equiv \mbox{Tr} \left(e^{- tH} \right)$ and the $S$-matrix:
\be
P(t) - P_0 (t) =  \int_{- \infty} ^{+ \infty} dE e^{-tE} \left[ \rho (E) - \rho_0 (E) \right]
\label{heatkernel}
\ee
and between the zeta function $\zeta_H (s) = \mbox{Tr} H^{-s}$ and the total phase shift:
\be
\zeta_H (s) - \zeta_{H_0} (s) = - \int_{- \infty} ^{+ \infty} dE E^{-s} {d \delta (E) \over dE}
\label{zeta}
\ee

We now introduce some terminologies used in different fields to describe related quantities.  From (\ref{smatrix}), we have that
\be
\mbox{det} S(z) = \mbox{det} {z - H \over z - H_0}
\ee
We can rewrite (\ref{thouless1}) in the form $\ln \mbox{det} \, S(z) = \mbox{Tr} \ln (z - H) - \mbox{Tr} \ln (z - H_0 )$. Then, using the definition of the counting function $\mathcal{N}(E)$, we have
\be
i \delta (z) = \int d\mathcal{N}(E) \ln (z -E) - \int d \mathcal{N}_0 (E) \ln (z -E)
\ee
The quantity $i \delta (z) = L(z)$ is sometimes called the Lyapunov function \cite{luck}. Taking $z = E \pm i \varepsilon$ with $\varepsilon \rightarrow 0^+$, we obtain
\be
\delta (E) = \pi \left( \mathcal{N}(E) - \mathcal{N}_0 (E) \right)
\ee
which states that the total phase shift measures the change of counting function up to energy $E$ in the presence of the scattering potential. The quantity $ \Delta (z) = e^{2 L(z)} = \mbox{det} (z- H) / (z - H_0)$  is often called the Fredholm determinant or the B\"ottcher function in the mathematics literature \cite{simon}.

There is an interesting relation established by Herbert and Jones and by Thouless \cite{thouless}, which identifies the real part of the Lyapunov function
\be
\gamma (E) \equiv \mbox{Re} L ( E + i \ep ) = \int dN(E') \ln |E - E' |
\label{loclength}
\ee
as the spatial decay (or localization) length of the corresponding energy state.

\subsection{Gelfand-Yaglom description and Transmission Probability}

In fact, in addition to the density of states $\Delta \rho (E)$, the Green's functions defined in the previous section \ref{sec:greens}, can also be used to compute the transmission probability  ${\mathcal T} = | t |^2$. Such a  combined description of spectral (density of states) and transport (total transmission probability) provides a very useful and powerful description of physical properties of quantum mesoscopic systems \cite{ambook,fisher-lee}.
The results can be summarized as follows.  Define the ratios of Green's functions to free Green's functions as:
$W_\pm = {G_\pm}/{G_\pm^{(0)}}$. Then
the transmission probability can be computed as  the {\it product} of  determinants:
\begin{eqnarray}
{\mathcal T}(E) = |t(E)|^2=\det \left(W_+(E)\right) \det \left(W_-(E)\right)  = \left | \det W_+(E) \right | ^2
%\mbox{det} S &=& {\det W_+ \over \det W_- }
\label{general}
\end{eqnarray}
while the density of states (\ref{dostates}) follows from the {\it  ratio} of  determinants:
\be
\Delta\rho(E)=\frac{1}{2\pi i} \frac{\partial}{\partial E}\, \ln\, \frac{\det\left(W_+(E)\right)}{\det\left(W_-(E)\right)}
\end{equation}
These relations  are valid in any space dimension, but are sometimes difficult to implement in specific problems since they involve formal definitions of determinants. However, for one-dimensional scattering problems these calculations simplify dramatically using the Gelfand-Yaglom theorem, which we now describe.

\section*{Gelfand-Yaglom theorem:}

At first sight one might think that computing the determinant of a differential operator is a hopeless task, as there are infinitely many eigenvalues, which then need to be multiplied together. However, the Gelfand-Yaglom theorem is a remarkable result that allows one to compute the determinant of a (one-dimensional) differential operator without computing any eigenvalues \cite{gy,levit,kleinert,dets}.

\begin{figure}[H]
%\centerline{\includegraphics[scale=0.4]{gelfand-yaglom.eps}}
\centerline{\includegraphics[scale=0.4]{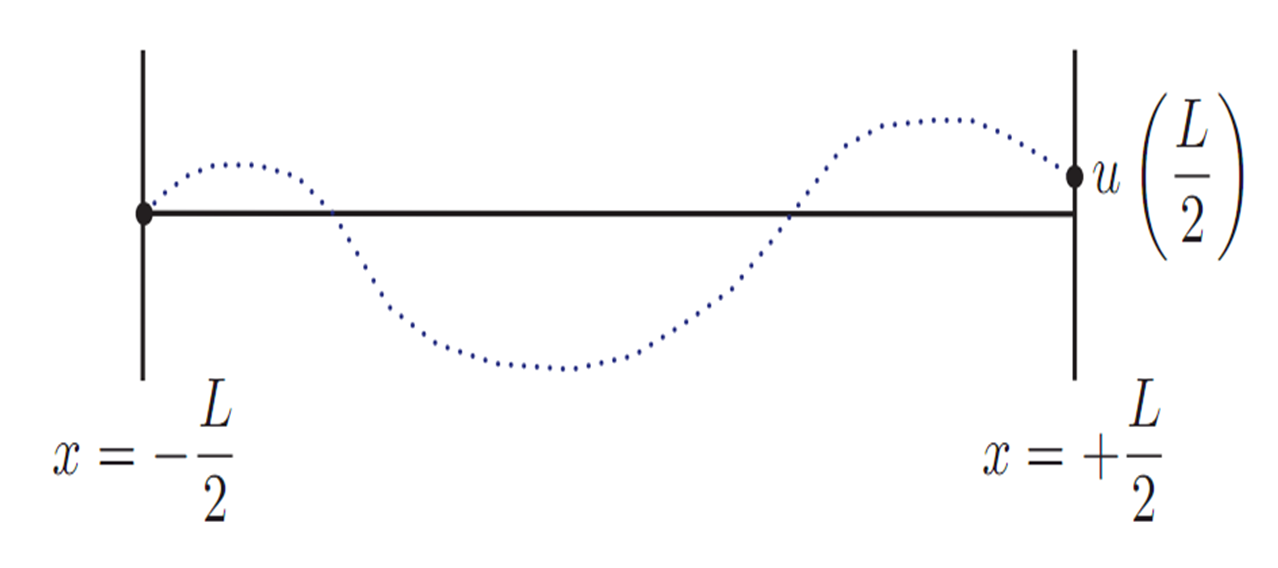}}
\caption{The Gelfand-Yaglom theorem \cite{gy,levit,kleinert,dets} states that to compute the determinant of a differential operator (with Dirichlet b.c.'s), one simply integrates numerically from one end of the interval to the other, with {\it  initial value conditions} that the function vanishes and has initial slope 1. Then the value of the function $u(L/2)$ at the other end of the interval is equal to the determinant. In fact, to be more precise this applies to {\it ratios} of determinants, as in (\ref{gy}).}
\label{GY}
\end{figure}

Consider the one-dimensional scattering problem with potential $V(x)$, as illustrated in the Figure \ref{GY} and the Hamiltonians  $H=-\frac{d^2}{dx^2}+V$, and $H_0=-\frac{d^2}{dx^2}$. According to (\ref{general}),  the transmission probability is
\begin{eqnarray}
{\mathcal T } (E) = |t(E)|^2=\left | \frac{\det\left(H_0-E-i\varepsilon\right)}{\det\left(H-E-i\varepsilon\right)} \right|^2 \, .
\label{tdet}
\end{eqnarray}
The determinant can be computed in an efficient way using the Gel'fand-Yaglom theorem, which for our purposes can be stated as follows:

Consider the Schr\"odinger eigenvalue problem $H\psi=\lambda\psi$, on the interval $x\in \left[-\frac{L}{2}, \frac{L}{2}\right]$, where $L$ is, as previously, much larger than the range of the potential $V(x)$. We apply  Dirichlet boundary conditions,  as this is relevant for the scattering problem, but the Gelfand-Yaglom theorem can be generalized to other boundary conditions \cite{dets}. Then the determinant of $H$ can be computed as follows: solve the {\it zero eigenvalue initial value problem}: $Hu=0$, with $u\left(-\frac{L}{2}\right)=0$ and $u^\prime\left(-\frac{L}{2}\right)=1$, and similarly for $H_0 u_0=0$. Then the Gelfand-Yaglom result \cite{gy,levit,kleinert,dets} states that
\begin{eqnarray}
\frac{\det H}{\det H_0}=\frac{u\left(\frac{L}{2}\right)}{u_0\left(\frac{L}{2}\right)}
\label{gy}
\end{eqnarray}
Notice that this means that all one needs to do is to numerically integrate, with simple initial value boundary conditions, from one end of the interval to another, which is straightforward to implement numerically, as illustrated in figure \ref{GY}. For other boundary conditions there are analogus results \cite{dets}.

To show how this is related to the transmission problem, we shift $H$ by $E$ , as in (\ref{tdet}), and write $E=k^2$. Then, near $x=-L/2$, where there is no potential, the solution of $(H-k^2)u=0$, satisfying the initial conditions, is
\begin{eqnarray}
u\sim \frac{\sin\left(k\left(x+\frac{L}{2}\right)\right)}{k}\sim\frac{1}{2i k}\left(e^{ik(x+L/2)}-e^{-i k (x+L/2)}\right)
\label{initial}
\end{eqnarray}
which has been decomposed into its "right-mover" and "left-mover" parts. Similarly, near $x=+L/2$, again the potential vanishes, so we can write
\begin{eqnarray}
u\sim A\, e^{ik(x+L/2)}+B\, e^{-i k (x+L/2)}
\label{final}
\end{eqnarray}
for some coefficients $A$ and $B$. To describe scattering from the right (for the probability, the direction does not matter \cite{konishi,visser}), we take $k\to k+i\varepsilon$, with $\varepsilon\to 0^+$, so that as $L\to \infty$ we select only the left-mover near $x=-L/2$. Near $x=+L/2$, this selects the left-mover, $e^{-i k (x+L/2)}$ term, which is the incident wave. Thus, the transmission amplitude is
\begin{eqnarray}
{\mathcal T} (k) =\frac{-\frac{1}{2ik} \, e^{-i k L/2}}{B\, e^{-i k L/2}}=-\frac{1}{2ik\, B}
\end{eqnarray}
But $u_0=\frac{1}{2i k}\left(e^{ik(x+L/2)}-e^{-i k (x+L/2)}\right)$, so that $u_0(L/2)=-\frac{1}{2i k}e^{-i k L}$ at $x=+L/2$. Similarly, $u(L/2)=B\, e^{-i k L}$. So:
\begin{eqnarray}
\frac{u(L/2)}{u_0(L/2)}=-2i\, k\, B=\frac{1}{{\mathcal T}}
\end{eqnarray}
from which the result (\ref{tdet}) follows.

%As an explicit example, consider the binding Eckart potential
%\begin{eqnarray}
%V=-j(j+1)\, {\rm sech}^2 x
%\label{eckart}
%\end{eqnarray}
%When $j$ is an integer this potential is reflectionless. But for any $j$ the Gel'fand-Yaglom determinant can be computed analytically [Dunne, lecture notes], since the Schr\"odinger equation is analytically soluble in terms of hypergeometric functions [Eckart, Morse/Feshbach]. One finds
%\begin{eqnarray}
%\frac{\det\left(-\frac{d^2}{dx^2}-j(j+1)\, {\rm sech}^2 x -E\right)}{\det\left(-\frac{d^2}{dx^2} -E\right)}=\frac{\Gamma(i\sqrt{E})\Gamma(1+i\sqrt{E})}{\Gamma(i\sqrt{E}-j)\Gamma(i\sqrt{E}+j+1)}
%\end{eqnarray}
%Figure ... shows a plot of a comparison of this function, for various values of $j$, compared with the Gelfand-Yaglom result obtained from a numerical integration using $L=10$, and also with a numerical computation from the Riccati equation (\ref{riccati2}). The curves agree perfectly.

%\section{Gelfand-Yaglom description}

\section{Transfer matrix formalism}
\label{sec:tmatrix}

In the previous section, we have advocated the description of wave propagation in complex systems using the $S$-matrix. It relates outgoing to incoming amplitudes in a unitary way, and provides a number of elegant expressions for spectral and transport quantities. However, in certain situations, such as for transport in one-dimensional problems, the {\it transfer matrix} formalism
offers computational advantages. For example, in layered systems it is useful to take advantage of the propagation of an incoming wave along a given fixed direction using the multiplicative property of the transfer matrix. To see how this works,  rewrite (\ref{smatrix0}) in an equivalent form, relating left-bound and right-bound amplitudes, instead of incoming and outgoing amplitudes (see figure \ref{fig1}):
\begin{eqnarray}
 \left(
\begin{array}{l}
i_R  \\
o_R
\end{array}
 \right)
 =
 \begin{pmatrix}
 {1 \over t}  & - {r \over t}   \cr
 {r^\prime \over t} & {1 \over t^*}
 \end{pmatrix}
% \left(
%\begin{array}{l|}
%{1 \over t}  &  {r \over t}   \\
% -{r^\prime \over t} & {1 \over t^*}
%\end{array}
% \right)
 \left(
\begin{array}{l}
o_L  \\
i_L
\end{array}
 \right)
 \equiv M
  \left(
\begin{array}{l}
o_L  \\
i_L
\end{array}
 \right)
\label{tmatrix}
\end{eqnarray}
which defines the transfer matrix $M$. Note that $\det M =1$. For a symmetric potential, namely for $r' = r$, $M$ takes the interesting symmetric form:
\begin{eqnarray}
M
 =   \left(
\begin{array}{ll}
1/t \, & \,  -r/t \\
- \left( r/t \right)^* \, & \, {1 / t^*}
\end{array}
 \right)  \label{tmatrixsym}
\end{eqnarray}

\medskip

{\bf Remark:} A great asset of the $S$-matrix and  $M$-matrix formulations is that they are very general and they do not require to specify a wave equation or any local properties such as group velocity, dispersion at least until some more specific calculations.

\medskip

The main advantage of the transfer matrix description (\ref{tmatrixsym}) over the $S$-matrix is that it allows to perform the calculation of the total transfer $M$ matrix  for the case of a one-dimensional periodic, quasi-periodic (fractal) case since we have a multiplication structure\cite{Born-wolf}. The transfer matrix can be viewed as a mapping transforming the wave after it passes through each scatterer or layer. Note first that the eigenvalues $\lambda_\pm$ of (\ref{tmatrixsym}) are solutions of $\lambda^2 - 2 \lambda \mbox{Re} \left( 1/t \right) + 1 =0$. Since $M$ is unimodular, $\lambda_+ \lambda_- =1$ and propagating wave solutions correspond to $-2 \leq 2 \mbox{Re} \left( 1/t \right) = \mbox{Tr} M \leq 2$, we can parameterize:
\be
\mbox{Tr} M \equiv 2 \cos \theta (k)
\label{theta}
\ee
which defines the so-called Bloch angle $\theta (k)$. The eigenvalues of $M(k)$ are thus $e^{\pm i \theta (k)}$.  There is an important relation between the Bloch angle and the total phase shift $\delta (k)$ defined in (\ref{dets}). To establish it, we start from the relation $ 1 - e^{ 2 i \delta} = 2 t \cos \theta$ between the two equivalent $S$ and $M$ formulations. Multiplying both sides by its complex conjugate, we obtain:
\be
\sin \delta (k) = \pm |t| \cos \theta (k) \, .
\label{blochs}
\ee
Using the expression (\ref{counting}), we obtain a relation between the counting function and the Bloch angle:
\be
\pi \Delta \mathcal{N}(k) = \delta (k) =  \mbox{Arcsin} \left( |t| \cos \theta (k) \right) \, .
\label{countingbloch}
\ee
This expression is useful numerically. It allows one to obtain directly the counting function and the density of modes by derivation from a counting of the accumulated Bloch angle. Before entering into more details for the case of the Helmholtz equation, we present now a slightly different but yet closely related point of view.

\section{Transfer matrices and a discrete Riccati equation}

\begin{figure}[H]
%\centerline{\includegraphics[scale=0.4]{step-potential.eps}}
\centerline{\includegraphics[scale=0.4]{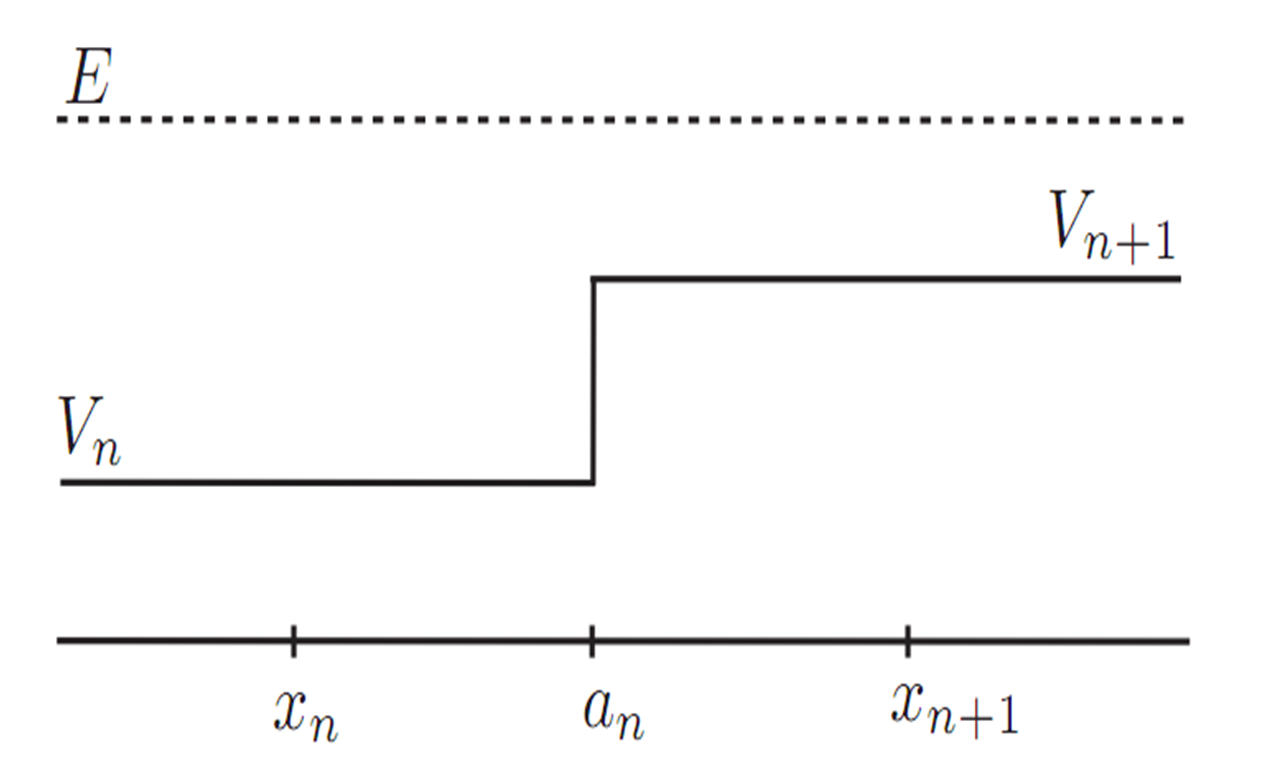}}
\caption{One segment of the discrete step potential for the discrete Riccati equation.}
\label{step}
\end{figure}

For convenience, we consider in this section the case of the Schr\"odinger equation. Consider scattering above a single step, as indicated in the figure \ref{step}. The wavefunctions to the left and right of the step, located at $x=a_n$, can be expressed as:
\begin{eqnarray}
\psi_n&=& \frac{1}{\sqrt{2k_n}}\left(A_n e^{i k_n x_n}+B_n e^{-i k_n x_n}\right)\quad, \quad x_n<a_n\\
\psi_{n+1}&=& \frac{1}{\sqrt{2k_{n+1}}}\left(A_{n+1}e^{i k_{n+1} x_{n+1}}+B_{n+1} e^{-i k_{n+1} x_{n+1}}\right)\quad, \quad x_{n+1}> a_n
\label{single-step}
\end{eqnarray}
where $k_n=\sqrt{E-V_n}$ and $k_{n+1}=\sqrt{E-V_{n+1}}$.

Matching $\psi$ and $\psi^\prime$ at $x=a_n$, we connect the coefficients of the left- and right-moving waves as:
\begin{eqnarray}
\begin{pmatrix}
A_{n+1} \cr B_{n+1}
\end{pmatrix}
=
\begin{pmatrix}
\alpha_n & \beta_n\cr
\beta_n^* & \alpha_n^*
\end{pmatrix}
\begin{pmatrix}
A_{n} \cr B_{n}
\end{pmatrix}
\label{transfer}
\end{eqnarray}
where
\begin{eqnarray}
\alpha_n=\frac{k_{n+1}+k_n}{2\sqrt{k_n k_{n+1}}}e^{i(k_n-k_{n+1})a_n}\quad, \quad
\beta_n=\frac{k_{n+1}-k_n}{2\sqrt{k_n k_{n+1}}}e^{-i(k_n+k_{n+1})a_n}
%\alpha_n&=&\frac{k_{n+1}+k_n}{2\sqrt{k_n k_{n+1}}}e^{i(k_n-k_{n+1})a_n}\\
%\beta_n&=&\frac{k_{n+1}-k_n}{2\sqrt{k_n k_{n+1}}}e^{-i(k_n+k_{n+1})a_n}
\end{eqnarray}
For scattering from the left we set $B_{n+1}=0$ and deduce the reflection and transmission amplitudes:
\begin{eqnarray}
r_n=\frac{B_n}{A_n}=-\frac{\beta_n^*}{\alpha_n^*}\quad, \quad
t_n=\frac{A_{n+1}}{A_n}=\frac{1}{\alpha_n^*}
%r_n&=&\frac{B_n}{A_n}=-\frac{\beta_n^*}{\alpha_n^*}\\
%t_n&=&\frac{A_{n+1}}{A_n}=\frac{1}{\alpha_n^*}
\end{eqnarray}
Conservation of probability is satisfied at each step, since $|r_n|^2+|t_n|^2=1$, which is equivalent to $|\alpha_n|^2-|\beta_n|^2=1$, and which corresponds to the fact that the transfer matrix defined in (\ref{transfer}) has unit determinant. The transfer matrix equation (\ref{transfer}) can also be expressed as an evolution equation for the $r_n$:
\begin{eqnarray}
r_{n+1}=\frac{\frac{\beta_n^*}{\alpha_n^*}+r_n}{\frac{\alpha_n}{\alpha_n^*}+\frac{\beta_n}{\alpha_n^*}\, r_n}
\label{discrete-riccati}
\end{eqnarray}
which we recognize as a discrete form of a Riccati equation.

\subsection{Continuous limit of discrete Riccati equation}

In order to take the continuous limit of a smoothly varying potential, for scattering above the potential, we think of the potential as broken into a series of discrete steps and successively apply the transfer matrix formalism of the previous section to pass from one step to the next. It proves convenient to use a slightly modified phase convention than in (\ref{single-step}): in the $x_n$ segment, $a_{n-1}<x_n <a_n$ we define
\begin{eqnarray}
\psi_n&=& \frac{1}{\sqrt{2k_n}}\left(e^{i\sum_j^{n-1}k_j(a_j-a_{j-1})}\tilde A_n e^{i k_n (x_n-a_{n-1})}+
e^{-i\sum_j^{n-1}k_j(a_j-a_{j-1})} \tilde B_n e^{-i k_n (x_n-a_{n-1})}\right)\nonumber \\
%\psi_{n+1}&=& \frac{1}{\sqrt{2k_{n+1}}}\left(A_{n+1}e^{i k_{n+1} x_{n+1}}+B_{n+1} e^{-i k_{n+1} x_{n+1}}\right)\quad, \quad x_{n+1}> a_n
\label{multiple-step1}
\end{eqnarray}
and in the $x_{n+1}$ segment, $a_{n}<x_{n+1} <a_{n+1}$ we define
\begin{eqnarray}
%\psi_&=& \frac{1}{\sqrt{2k_n}}\left(e^{i\sum_j^{n-1}k_j(a_j-a_{j-1})}A_n e^{i k_n (x_n-a_{n-1})}+
%e^{-i\sum_j^{n-1}k_j(a_j-a_{j-1})} B_n e^{-i k_n (x_n-a_{n-1})}\right)
\psi_{n+1}&=& \frac{1}{\sqrt{2k_{n+1}}}\left(e^{i\sum_j^{n}k_j(a_j-a_{j-1})} \tilde A_{n+1}e^{i k_{n+1} (x_{n+1}-a_n)} \right. \nonumber\\
&& \hskip 2cm \left. +
e^{-i\sum_j^{n}k_j(a_j-a_{j-1})} \tilde B_{n+1} e^{-i k_{n+1} (x_{n+1}-a_n)}\right)
%\psi_{n+1}&=& \frac{1}{\sqrt{2k_{n+1}}}\left(e^{i\sum_j^{n}k_j(a_j-a_{j-1})} \tilde A_{n+1}e^{i k_{n+1} (x_{n+1}-a_n)}+
%e^{-i\sum_j^{n}k_j(a_j-a_{j-1})} \tilde B_{n+1} e^{-i k_{n+1} (x_{n+1}-a_n)}\right) \nonumber\\
\label{multiple-step2}
\end{eqnarray}
This phase convention refers the phase to the left-hand end of the segment, and keeps a total accumulated phase from previous segments. Then the transfer matrix relation (\ref{transfer}) becomes:
\begin{eqnarray}
\begin{pmatrix}
\tilde A_{n+1} \cr \tilde B_{n+1}
\end{pmatrix}
=
\begin{pmatrix}
\tilde \alpha_n & \tilde \beta_n\cr
\tilde \beta_n^* & \tilde \alpha_n^*
\end{pmatrix}
\begin{pmatrix}
\tilde A_{n} \cr \tilde B_{n}
\end{pmatrix}
\label{transfer1}
\end{eqnarray}
where now $\tilde{\alpha}_n$ is real:
\begin{eqnarray}
\tilde \alpha_n = \frac{k_{n+1}+k_n}{2\sqrt{k_n k_{n+1}}}\quad, \quad
\tilde \beta_n = \frac{k_{n+1}-k_n}{2\sqrt{k_n k_{n+1}}}e^{-2i\sum_j^n k_j(a_j-a_{j-1})}
%\tilde \alpha_n&=&\frac{k_{n+1}+k_n}{2\sqrt{k_n k_{n+1}}}\\
%\tilde \beta_n&=&\frac{k_{n+1}-k_n}{2\sqrt{k_n k_{n+1}}}e^{-2i\sum_j^n k_j(a_j-a_{j-1})}
\label{alphabeta2}
\end{eqnarray}
Clearly we still have unitarity: $|\tilde\alpha_n|^2-|\tilde \beta_n|^2=1$, for all $n$.  Defining a reflection amplitude at step $n$ as $\tilde r_n\equiv \tilde B_n/\tilde A_n$, the evolution equation (\ref{discrete-riccati}) can be written as
\begin{eqnarray}
\tilde r_{n+1}-\tilde r_n=\frac{\frac{\tilde \beta_n^*}{\tilde \alpha_n^*}+\frac{(\tilde\alpha_n^*-\tilde\alpha_n)}{\tilde\alpha_n^*}\,\tilde r_n-\frac{\tilde \beta_n}{\tilde \alpha_n^*}\, \tilde r_n^2}{\frac{\tilde \alpha_n}{\tilde \alpha_n^*}+\frac{\tilde \beta_n}{\tilde \alpha_n^*}\, \tilde r_n}
=\frac{\tilde\beta_n^*}{\tilde\alpha_n}\left(\frac{1-\frac{\tilde\beta_n}{\tilde\beta_n^*}\,\tilde r_n^2}{1+\frac{\tilde\beta_n}{\tilde\alpha_n}\,\tilde r_n}\right)
\label{discrete-riccati2}
\end{eqnarray}
where we have used the fact that the $\tilde\alpha_n$ are real with this phase convention.

To take the continuous limit, we take the $a_n$ to be equally spaced by distance $\Delta$, and note that
\begin{eqnarray}
\frac{k_{n+1}-k_n}{\Delta} &\to& -\frac{1}{2}\frac{1}{\Delta}\frac{V_{n+1}-V_n}{\sqrt{E-V_n}}\to -\frac{1}{2} \frac{V^\prime(x)}{Q(x)}\\
\frac{1}{k_{n+1}+k_n}&\to& \frac{1}{2Q(x)}
\end{eqnarray}
where $Q(x)=\sqrt{E-V(x)}$. Therefore,
\begin{eqnarray}
\frac{1}{\Delta}\frac{\tilde\beta^*_n}{\tilde\alpha_n^*}&=&\frac{1}{\Delta}\left(\frac{k_{n+1}-k_n}{k_{n+1}+k_n}\right) e^{2i\sum_j^n k_j(a_j-a_{j-1})}\to \frac{Q^\prime(x)}{2Q(x)}\, e^{2i\int^x Q(x^\prime)\, dx^\prime}
\end{eqnarray}
and similarly for the conjugate. Furthermore,  $\tilde\beta_n/\tilde\alpha_n^*=O(\Delta)\to 0$. Therefore, the continuum limit of the discrete Riccati equation (\ref{discrete-riccati2}) is
\begin{eqnarray}
\tilde r^\prime(x)=\frac{Q^\prime}{2Q}\left(e^{2i\int^x Q}-\tilde r^2(x) e^{-2i\int^x Q}\right)
\label{riccati}
\end{eqnarray}
This is straightforward to implement numerically:  integrate for scattering from the left with the initial condition $\tilde r(-\infty)=0$, yielding the reflection amplitude as $\tilde r(+\infty)$.

For comparison we briefly recall the derivation of this continuous Riccati equation (\ref{riccati}) for one-dimensional scattering \cite{berry-mount}. Re-write the wavefunction $\psi(x)$ in terms of WKB adiabatic solutions with coefficients $\tilde A(x)$ and $\tilde B(x)$ as:
\begin{eqnarray}
\psi(x)&=&\frac{1}{\sqrt{2Q(x)}}\left(\tilde A(x)e^{i\int^x Q}+\tilde B(x) e^{-i\int^x Q}\right)
\\
\psi^\prime(x)&=&\frac{i Q(x)}{\sqrt{2Q(x)}}\left(\tilde A(x)e^{i\int^x Q}-\tilde B(x) e^{-i\int^x Q}\right)
\label{decomp}
\end{eqnarray}
where, as above, $Q(x)=\sqrt{E-V(x)}$. The decomposition (\ref{decomp})
requires the following relation between the (complex) coefficient functions:
\begin{eqnarray}
\begin{pmatrix}
\tilde A^\prime(x)\cr \tilde B^\prime(x)
\end{pmatrix}
=
\frac{Q^\prime(x)}{2Q(x)}
\begin{pmatrix}
0 & e^{-2i\int^x Q} \cr
e^{2i\int^x Q} & 0
\end{pmatrix}
\begin{pmatrix}
\tilde A(x)\cr \tilde B(x)
\end{pmatrix}
\label{transfer-contin}
\end{eqnarray}
Then defining the reflection coefficient amplitude $\tilde r(x)=\frac{\tilde B(x)}{\tilde A(x)}$, we see that it satisfies the differential equation
\begin{eqnarray}
\tilde r^\prime=\frac{\tilde A\tilde B^\prime-\tilde B \tilde A^\prime}{\tilde A^2}=
\frac{Q^\prime}{2Q}\left(e^{2i\int^x Q}-\tilde r^2 e^{-2i\int^x Q}\right)
\label{riccati2}
\end{eqnarray}
in agreement with the limit (\ref{riccati}) of the discrete transfer matrix approach.
\\

{\bf Comments:}
\begin{enumerate}

\item
We should distinguish the Riccati form (\ref{riccati2}) from another, perhaps more familiar, realization of the
Schr\"odinger equation as a Riccati equation: define a "momentum" function in terms of the derivative of the wavefunction $\psi(x)$ as $p=-i\frac{d}{dx}\ln \psi$, and the Schr\"odinger equation for $\psi(x)$ becomes a Riccati equation for $p(x)$
\begin{eqnarray}
p^\prime=iQ-i\,p^2
\label{riccati3}
\end{eqnarray}
A simple computation shows that the two Riccati equations (\ref{riccati2}) and (\ref{riccati3}) are completely equivalent, with the functions $p(x)$ and $\tilde r(x)$  related by:
\begin{eqnarray}
p=-Q\left(\frac{1-\tilde r\, e^{2 i\int^x Q}}{1+\tilde r \, e^{2 i\int^x Q}}\right)\qquad, \qquad \tilde r=\, e^{-2 i\int^x Q}\left(\frac{Q+p}{Q-p}\right)
\label{pr}
\end{eqnarray}

\item
It is worth noting that when integrating a Riccati differential equation, such as $y^\prime(x)=b(x)-c(x)\, y^2(x)$,  it often proves numerically useful to use a M\"obius integrator  \cite{schiff} for step $\Delta$
\begin{eqnarray}
y_{n+1}=\frac{b_n\, \Delta +y_n}{1+c_n \, y_n\, \Delta}\qquad ,
\label{mobius}
\end{eqnarray}
rather than a conventional Numerov integrator. Clearly, the discrete Riccati equation (\ref{discrete-riccati}) is precisely of this form, when we make the phase choice  (\ref{alphabeta2}).

\end{enumerate}

\section{Transfer matrix description for the Helmholtz equation}

%%%%%%%%%%%%%%%%%%%%%%%%%%%%%%%%%%%%%%%%%%%%%%%%%%
%START OF ELI'S PART

The purpose of this section is to implement the previous methods in the case of the propagation of  electromagnetic waves in one-dimensional (1D) waveguide structures built out of complex media. We first reiterate the basic equations and then consider specific cases of layered dielectric media : free space, Fabry-Perot structure, periodic structures (photonic crystals), random systems and single impurity. Finally, we discuss in more detail the case of a Fibonacci potential. In all these cases, we obtain spectral quantities (transmission, counting function and density of modes) and the steady state (stationary) local structure of the electric field.

\medskip

The use of the transfer matrix method for solving Maxwell
equations in photonic systems essentially converts them into a set
of fi{}nite difference equations in real space and then rearranges
those equations into the form of a transfer matrix \cite{Born-wolf}.
In 1D structures, transfer matrices relate the electric and magnetic
fi{}elds in one plane those in an adjacent plane. The method
can serve as an approximation for any dielectric stratified medium
with $\epsilon=\epsilon\left(z\right)\:,\mu=\mu\left(z\right)$.
In the case of an arbitrary finite structure of piecewise constant
potential, the transfer matrix method is an exact solution.

\begin{figure}[ht]
\centerline{\includegraphics[scale=0.45]{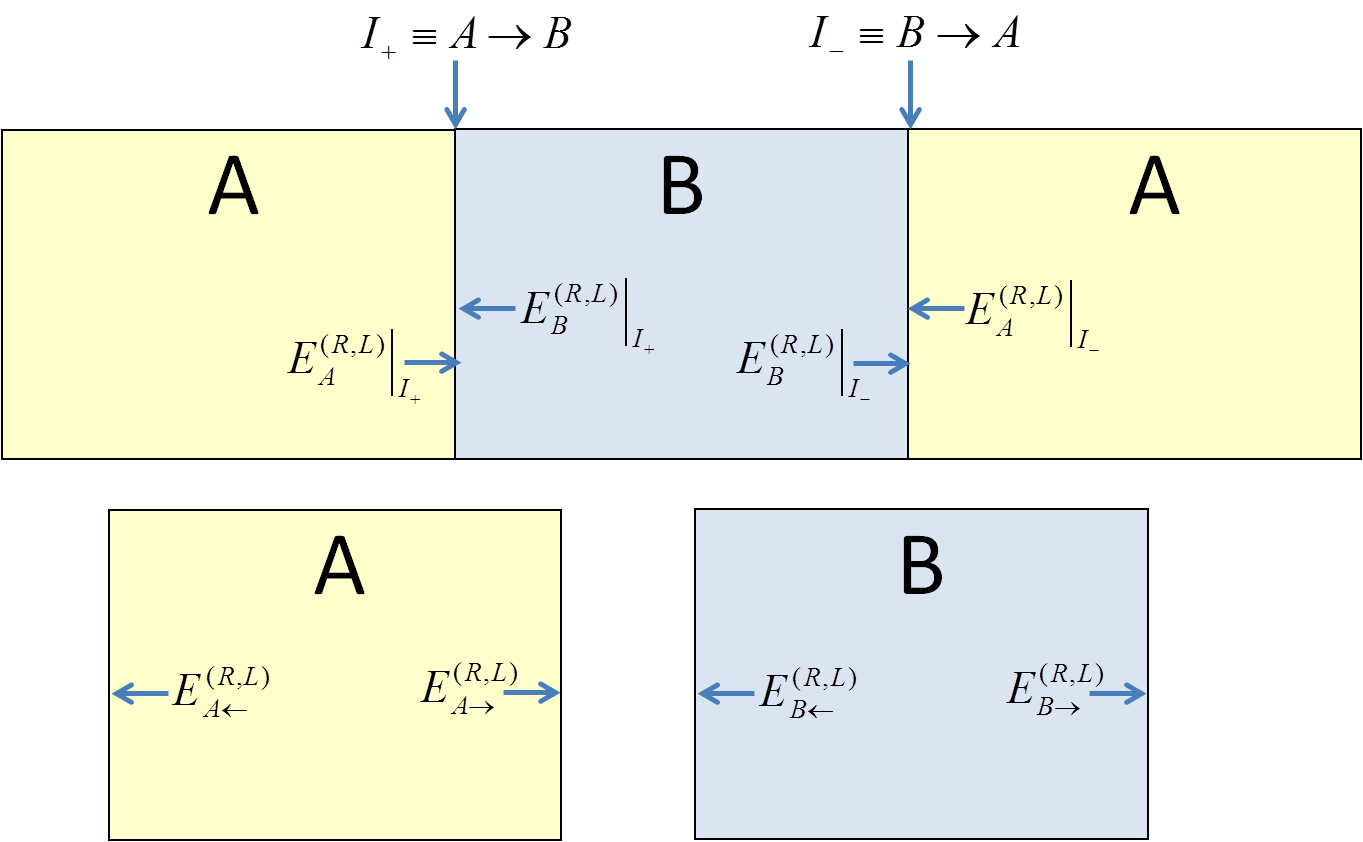}}
\caption{Setup of notations for the continuity relations at an interface ABA (upper figure) and for the free propagation through a slab (lower figure). Upper part: fields at the interfaces AB and BA are noted. Lower part: fields at the right and left ends of slab A and B are noted.}
\label{figIntNot}
\end{figure}

\begin{figure}[ht]
\includegraphics[scale=0.30]{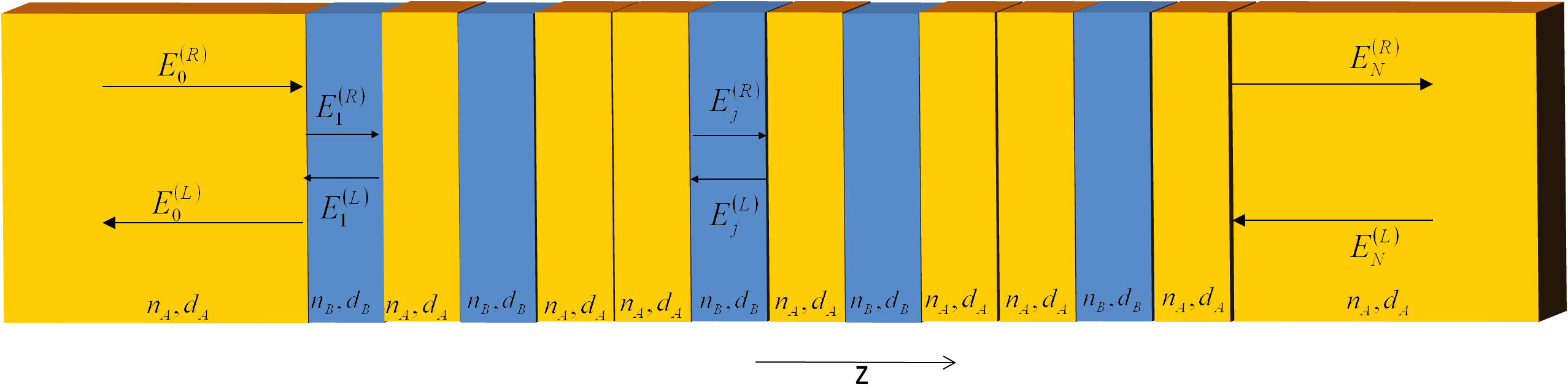}
\caption{Setup of notation for the 1D transfer matrix model for a binary layered structure. Here $z$ is the $1D$ coordinate, $d_{A,B}$ is the slab thickness, $n_{A,B}$ is the refractive index, and $E^{R,L}$ is the right-bound and left-bound electric field amplitudes.}
\label{fignotation}
\end{figure}

The general solution for the transverse electric (TE)  electric field $E_x$ propagating along the $z$-direction in a 1D structure is obtained from the solution of the wave equation
$c^2 \partial_{z}^{2}E_{x} - \partial_{t}^{2}E_{x}=0$. It takes the form
\be
E_{x}\left(z,t\right)=E_{x}^{(R)}\left(z_{0}\right)e^{i\left(k_{z}\left(z\right)z-\omega t\right)}+E_{x}^{(L)}\left(z_{0}\right)e^{i\left(-k_{z}\left(z\right)z-\omega t\right)}
\label{tefield}
\ee
where
$E_{x}^{(R)}\left(z_{0}\right)$ and $E_{x}^{(L)}\left(z_{0}\right)$ denote rightbound $(R)$ and leftbound $(L)$
waves steady-state amplitudes evaluated at the origin $z_{0}$ of the structure. Setting $k (z)=n\left(z\right)k$,
we can write for a monochromatic wave of frequency $\omega$,
\be
E_{x}\left(z\right)=E_{x}^{(R)}\left(z_{0}\right)e^{i k(z) z}+E_{x}^{(L)}\left(z_{0}\right)e^{-i k(z) z} \, .
\label{Exzstat}
\ee
 Using the definition (\ref{tmatrix}) of the transfer matrix M, we obtain
 \be
 \left(\begin{array}{c}
E_{x}^{(R)}\\
E_{x}^{(L)}
\end{array}\right)_{z_{end}}\equiv M\left(\begin{array}{c}
E_{x}^{(R)}\\
E_{x}^{(L)}
\end{array}\right)_{z_{0}}
\label{matrix91}
\ee
where $z_{0}$ and $z_{end}$ are the coordinates of the two outer boundaries
of the structure. In
non-absorbing dielectric structures considered here, conservation of energy implies that M is a unitary $2 \times 2$ complex valued matrix of unity determinant. In a layered structure $z_0 < z_1 < \cdots < z_{end}$, and denoting $M_{{z_i} \rightarrow {z_{i+1}}}$ the transfer matrix associated to a layer, we have a general multiplication scheme:
\be
M_{z_{0}\rightarrow z_{end}}=M_{z_{end-1}\rightarrow z_{end}} \, \cdots \, M_{z_{1}\rightarrow z_{2}} \,  M_{z_{0}\rightarrow z_{1}} \, .
\ee
For practical purposes we consider from now on, the scheme represented in Figures \ref{figIntNot} and \ref{fignotation} is a bit more intricate. Namely, each $M_{{z_i} \rightarrow {z_{i+1}}}$ appears as a product of sub-matrices responsible either for an interface boundary between two adjacent dielectrics or for the free propagation through a homogeneous dielectric slab.

%In the case of a periodic structure, the transfer matrix can be constructed
%from the sub transfer matrix of the unit cell:

%For the case of 1D structures of aperiodic order, the refraction index
%along the z axis falls into the category of a piecewise constant potential,
%and there for the transfer matrix description is exact. We present
%here a transfer matrix numerical algorythm for any finite binary 1D
%dielectric structures following the work of Kohmoto \cite{Kohmotocanonic},
%which can be easily writen in computer code. The mathematical notation
%we use is depicted in the following figure:

We will now restrict ourselves to  the more specific case of a binary layered structure built out of  $N$ slabs of two types of dielectrics $A$ and $B$,
where $n_{A,B}$ are the respective refractive indices in type $A$ and $B$ slabs of respective corresponding thicknesses $d_{A,B}$. The notations for this case are given in figures \ref{figIntNot} and \ref{fignotation}. The solution (\ref{Exzstat}) for the electric field in a single slab of type  $A$ or $B$ can be written as:
% $j=1,2...N$ is the slab number
%(N is the total number of slabs), z is the coordinate, and E is the
%electric field amplitude. The superscript (1) and (2), as before indicates
%rightbound and leftbound plan waves respectively. The general form
%for the TE electric amplitude in a slab of type A(B) in a 1D dielectric
%stack is:

\be
E_{A,B} = E_{A,B}^{(R)} \, e^{i(n_{A,B} \, k \, z)}+E_{A,B}^{(L)} \, e^{i(-n_{A,B}\, k \, z)}
\label {slabef}
\ee
where $E_{A,B}^{(R)}$ and $E_{A,B}^{(L)}$ are the right-bound and left-bound electric fields evaluated at the left boundary of slab $A,B$.

\medskip

{\bf Remark:} As stated in the Remark in section \ref{sec:tmatrix}, the $M$ (or $S$)-matrix formalism does not require precise knowledge of a local wave equation inside the structure and can be formulated without the input of a local group velocity (even ill-defined in the present case due to the discontinuity of the refractive index in the layered structure). But at some point the writing of relations (\ref{Exzstat}) and (\ref{slabef}) rely on  the definition of a wave vector and a group velocity : here we assume free propagation in each individual slab.
\medskip

An interface $I \equiv A\leftrightarrow B$ between adjacent slabs $A$ and $B$ is defined in figure \ref{figIntNot}.  The boundary conditions at a dielectric interface are imposed through continuity relations:
\be
%\begin{cases}
%\begin{array}{c}
%\left.
E_{A}^{(R)}+E_{A}^{(L)} |_{I_{+,-}} = E_{B}^{(R)}+E_{B}^{(L)} |_{I_{+,-}} \, \, \, \,  (\mbox{continuity})
\label{cont1}
\ee
%\left.
and
\be
n_{A} \left(E_{A}^{(R)} - E_{A}^{(L)}\right) |_{I_{+,-}} = n_{B} \left(E_{B}^{(R)}-E_{B}^{(L)}\right) |_{I_{+,-}} \, \, \, (\mbox{derivative-continuity})
%\end{array}\end{cases}
\label{continuity}
\ee
The phase accumulation which accounts for free propagation through a slab $A$ or $B$ is:
\be
\begin{cases}
\begin{array}{c}
E_{A\rightarrow,B\rightarrow}^{(R)}=E_{A\leftarrow,B\leftarrow}^{(R)}e^{in_{A,B}d_{A,B}k}\equiv E_{A\leftarrow,B\leftarrow}^{(R)}e^{i\delta_{A,B}}\\
E_{A\rightarrow,B\rightarrow}^{(L)}=E_{A\leftarrow,B\leftarrow}^{(L)}e^{-in_{A,B}d_{A,B}k}\equiv E_{A\leftarrow,B\leftarrow}^{(L)}e^{-i\delta_{A,B}}\\
\end{array}\end{cases}
\label{phase accumulation}
\ee
where we have defined,
\be
\delta_A\equiv n_Ad_Ak \qquad, \qquad  \delta_B\equiv n_Bd_Bk \, .
\label{deltadef}
\ee
It is convenient to set the layer thickness so that $n_{A}d_{A}=n_{B}d_{B}$ which leads to $\delta_A=\delta_B$. With the change of variables:
\be
\begin{cases}
\begin{array}{c}
E_{A,B}^{(+)}\equiv E_{A,B}^{(R)}+E_{A,B}^{(L)}\:\:\:\:\\
E_{A,B}^{(-)}\equiv-i\left(E_{A,B}^{(R)}-E_{A,B}^{(L)}\right) \, ,
\end{array}\end{cases}
\label{e+e-}
\ee
continuity equations (\ref{cont1}) and (\ref{continuity}) can be rewritten as:

\be
\protect\begin{cases}
\protect\begin{array}{c}
\left.E_{A,B}^{(+)}\right|_{I_{+,-}}\:=\left.E_{B,A}^{(+)}\right|_{I_{+,-}}\:\:\:\protect\\
\left.E_{A,B}^{(-)}\right|_{I_{+,-}}=\frac{n_{B,A}}{n_{A,B}}\left.E_{B,A}^{(-)}\right|_{I_{+,-}}
\protect\end{array}\protect\end{cases}
\ee
This leads to the definition of  four transfer matrices defined by the two sets of relations,
\be
%\begin{cases}
\begin{array}{c}
\left.\left(\begin{array}{c}
E^{(+)}\\
E^{(-)}
\end{array}\right)_{A}\right|_{I_-}\equiv T_{AB}\left.\left(\begin{array}{c}
E^{(+)}\\
E^{(-)}
\end{array}\right)_{B}\right|_{I_-}\:\\
\left.\left(\begin{array}{c}
E^{(+)}\\
E^{(-)}
\end{array}\right)_{B}\right|_{I_+}\equiv T_{BA}\left.\left(\begin{array}{c}
E^{(+)}\\
E^{(-)}
\end{array}\right)_{A}\right|_{I_+}
%\:\\
\end{array}
\label{matset1}
\ee
and
\be
\begin{array}{c}
\left(\begin{array}{c}
E^{(+)}\\
E^{(-)}
\end{array}\right)_{A\rightarrow}\equiv T_{A}\left(\begin{array}{c}
E^{(+)}\\
E^{(-)}
\end{array}\right)_{A\leftarrow}\\
\left(\begin{array}{c}
E^{(+)}\\
E^{(-)}
\end{array}\right)_{B\rightarrow}\equiv T_{B}\left(\begin{array}{c}
E^{(+)}\\
E^{(-)}
\end{array}\right)_{B\leftarrow}
\end{array}
%\end{cases}
\label{matset2}
\ee
With the aid of equations (\ref{deltadef}) and (\ref{e+e-}), we obtain four real valued matrices \cite{quasi,kohomotoothers}:
\be
%\begin{cases}
%\begin{array}{c}
T_{AB}=\left(\begin{array}{cc}
1 & 0\\
0 & \frac{n_{B}}{n_{A}}
\end{array}\right)=T_{BA}^{-1} \, \, \,  ,
%\mbox{and}
\, \, \,
%\:\:\\
T_{BA}=\left(\begin{array}{cc}
1 & 0\\
0 & \frac{n_{A}}{n_{B}}
\end{array}\right)=T_{AB}^{-1}
%\:\:\\
\label{tabeq1}
\ee
and
\be
T_{A,B}=\left(\begin{array}{cc}
cos \, \delta_{A,B} & -sin \, \delta_{A,B}\\
sin \, \delta_{A,B} & cos \, \delta_{A,B}
\end{array}\right)
% \, \, \, ,
%\mbox{and}
%\, \, \,
%T_{B}=\left(\begin{array}{cc}
%cos\delta_{B} & -sin\delta_{B}\\
%sin\delta_{B} & cos\delta_{B}
%\end{array}\right)
%\end{array}
%\end{cases} \, .
\label{submatresult}
\ee
so that the total transfer matrix $M_{N}$ of N dielectric slabs
is
\be
\left(\begin{array}{c}
E^{(+)}\\
E^{(-)}
\end{array}\right)_{N}\equiv M_{N}\left(\begin{array}{c}
E^{(+)}\\
E^{(-)}
\end{array}\right)_{0} \, .
\ee
To illustrate the efficiency of this systematic approach to the calculation of $M_{N}$ as a product
of an appropriate sequence of the four sub-matrices (\ref{submatresult}), we consider the matrix $M_{8}$ describing an arbitrary structure inserted  between slabs of type A (to preserve unitarity) as:
\be
M_{8}({[A]}{[ABABAABA]}{[A]})=T_{A}T_{AB}T_{B}T_{BA}T_{A}T_{A}T_{AB}T_{B}T_{BA}T_{A}T_{AB}T_{B}T_{BA}
\ee
Once the transfer matrix is obtained, an algebraic transformation between (\ref{smatrix0}) and (\ref{tmatrix}) allows
to derive the scattering matrix
\be
\left(\begin{array}{c}
E_{0}^{(R)}\\
E_{N}^{(L)}
\end{array}\right)\equiv S_{N}\left(\begin{array}{c}
E_{0}^{(L)}\\
E_{N}^{(R)}
\end{array}\right)
\ee
which relates incoming  to outgoing fields. From these expressions, we deduce the transmission
and reflection coefficients $\mathcal{T}\left(\omega\right)$ and $\mathcal{R}\left(\omega\right)$, the density of modes $\rho\left(\omega\right)$ and the counting function $\mathcal{N}\left(\omega\right)$.

%%%%%%HERE%%%%%%%%%%

\paragraph{Transmission and reflection spectra :}

The transmission and reflection spectra are calculated  for the case of an incident wave
from  the left (see \ref{amplitudedef})
%Generally, this calculation can be performed for the cases of irradiation from the right, or irradiation from both sides. Additionally, all
and spectra are normalized by the incoming plane wave electric field
amplitude, namely,
\be
\left\{ \begin{array}{c}
E_{0}^{(R)}=1\\
E_{N}^{(R)}=t(\omega)\\
E_{0}^{(L)}=r(\omega)\\
E_{N}^{(L)}=0
\end{array}\quad;\quad\left\{ \begin{array}{c}
\mathcal{T}(\omega)=\left|t(\omega)\right|^{2}\\
\mathcal{R}(\omega)=\left|r(\omega)\right|^{2}
\end{array}\right.\right.
\label{amplitudedef}
\ee
Using the transfer matrix $M_{tot}$, defined
by $\left(\begin{array}{c}
t\\
\frac{t}{i}
\end{array}\right)\equiv M_{tot}\left(\begin{array}{c}
1+r\\
\frac{1-r}{i}
\end{array}\right)$, yields:

\be
\begin{array}{c}
\mathcal{T}\left(\omega\right)=\frac{4\left|M\left(\omega\right)\right|^{2}}{\left\Vert M\left(\omega\right)\right\Vert ^{2}+2}=\frac{4}{\left\Vert M\left(\omega\right)\right\Vert ^{2}+2}\quad\quad\quad\quad\quad\quad\quad\quad\\
\begin{array}{c}
\\
\mathcal{R}\left(\omega\right)=\frac{\left\Vert M\left(\omega\right)\right\Vert ^{2}-2\left|M\left(\omega\right)\right|}{\left\Vert M\left(\omega\right)\right\Vert ^{2}+2\left|M\left(\omega\right)\right|}=\frac{\left\Vert M\left(\omega\right)\right\Vert ^{2}-2}{\left\Vert M\left(\omega\right)\right\Vert ^{2}+2}=1-\mathcal{T}\left(\omega\right)
\end{array}
\end{array}
\ee
where ${\left\Vert M\right\Vert^{2}}\equiv M_{11}^2+M_{12}^2+M_{21}^2+M_{22}^2$, and $M = [M_{ij}]$.

\paragraph{Density of modes spectrum calculation:}

The density of modes (DOM), $\rho\left(\omega\right)$, can now be calculated from either the scattering or transfer matrix, utilizing the total phase shift $\delta(\omega)$ or the phase of the transmitted amplitude $\alpha\left(\omega\right)$ defined in section (\ref{sec:smatrix}). For a 1D system of length L, the normalized DOM given by (\ref{theequation}) is $\rho\left(\omega\right)\equiv \frac{1}{L}\frac{d\alpha\left(\omega\right)}{d\omega}= \frac{1}{L}\frac{d\delta\left(\omega\right)}{d\omega}$. The normalization with the length $L$ is convenient as it keeps the DOM of a free system equal to one, and simplifies the evaluation of the relative enhancement of $\rho\left(\omega\right)$ for a structured system compared to a free one. This normalization also allows, in the case of aperiodic inhomogeneous systems, to give a meaning to otherwise ill defined quantities such as the effective group velocity and wave vector.
The calculation of $\rho\left(\omega\right)$ using either  scattering (\ref{theequation}) or transfer\cite{tmatrixrho} matrices is thus based on,
\begin{eqnarray}
\left\{ \begin{array}{c}
\begin{array}{c}
\rho\left(\omega\right)={1 \over L}\, \mbox{Im} \,\frac{\partial}{\partial \omega} \ln \mbox{det}\, S(\omega)
\\
\\
\end{array}\\
\rho\left(\omega\right)= \frac{1}{L}\frac{y'\left(\omega\right)x\left(\omega\right)-x'\left(\omega\right)y\left(\omega\right)}{x^{2}\left(\omega\right)+y^{2}\left(\omega\right)}\quad;\quad t\left(\omega\right)=x\left(\omega\right)+iy\left(\omega\right)
\end{array}\right.
\label{rhoofomega}
\end{eqnarray}
%For more detail regarding the scattering and transfer matrix calculation of $\rho\left(\omega\right)$ refer to ?? and ??.

\paragraph*{Amplitude of the local electric field :}

The electric field amplitude and intensity $\omega-z$ maps are obtained by calculating $E\left(\omega\right)$
for each $z$ coordinate inside the structure, using
two transfer matrices: the total transfer matrix $M_{tot}$, and the
partial transfer matrix from the left end to the point of interest
$M_{j}$. Firstly, $r$ and $t$ are calculated as before using $M_{tot}$,
and then we can use $M_{j}$ to calculate the field and intensity
in slab $j$ using:

\be
\begin{cases}
\begin{array}{c}
\left(\begin{array}{c}
E^{(+)}\\
E^{(-)}
\end{array}\right)_{j}=M_{j}\left(\begin{array}{c}
1+r\\
\frac{1-r}{i}
\end{array}\right)\\
E_{j}^{net}=Re\left[E_{j}^{\left(+\right)}\right]\quad\quad\\
I_{j}^{net}=\left(E_{j}^{net}\right)^{2}\quad\quad\quad
\end{array}\end{cases}
\ee
The net electric amplitude $E(z)^{net}$ at a point $z$, is the coherent amplitude sum of multiply reflected right-bound and left-bound waves, incorporating interference effects inside the system.
This process is repeated for all $j$ in order to create the full electric
field $E^{net}\left(\omega,z\right)$ and intensity distribution $I^{net}\left(\omega,z \right)$ maps. The spatial resolution can be enhanced by numerical subdivisions of the slabs. This calculation can be performed for the cases of an incident wave either from left, right  or from both sides.

\section{Illustrative Examples of Layered Systems}

\subsection{Free space}

\begin{figure}[H]
\centerline{\includegraphics[scale=0.30]{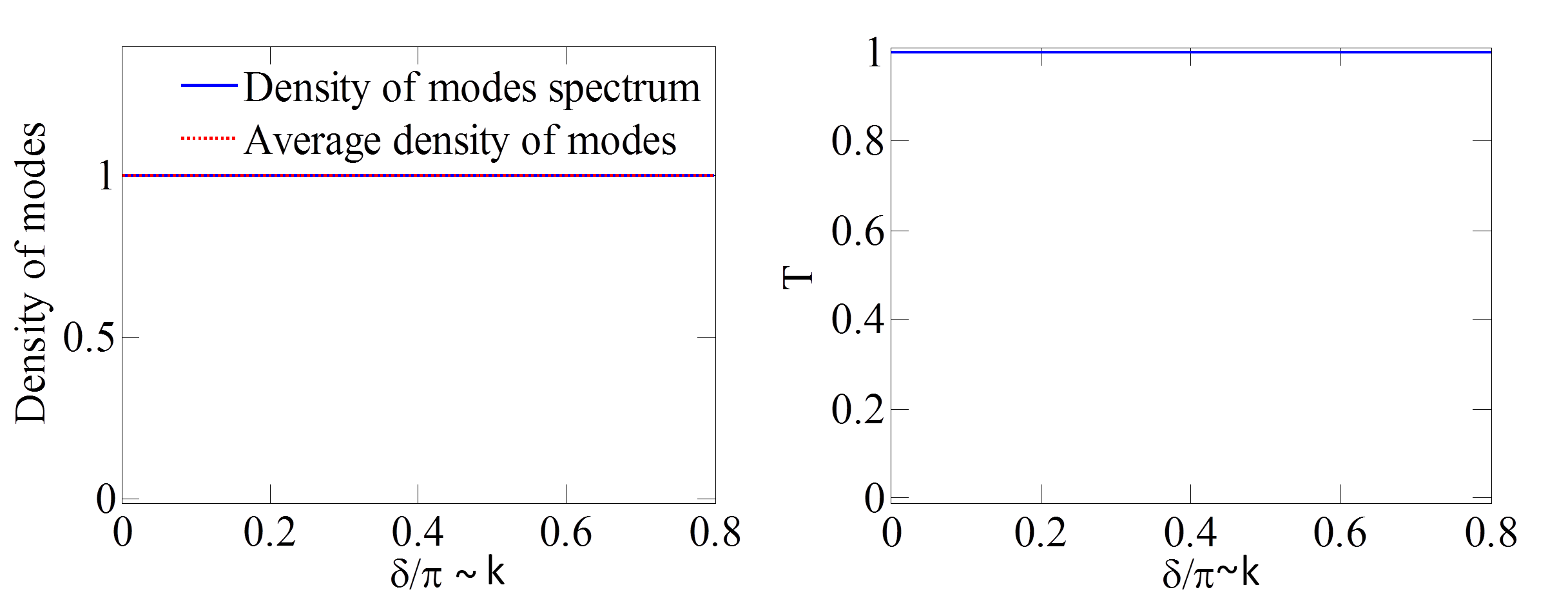}}
\caption{The normalized density of modes (left part) and the transmission spectrum (right part),  for free space (no scattering structure).}
\label{figTRhofreespace}
\end{figure}

\begin{figure}[H]
\centerline{\includegraphics[scale=0.30]{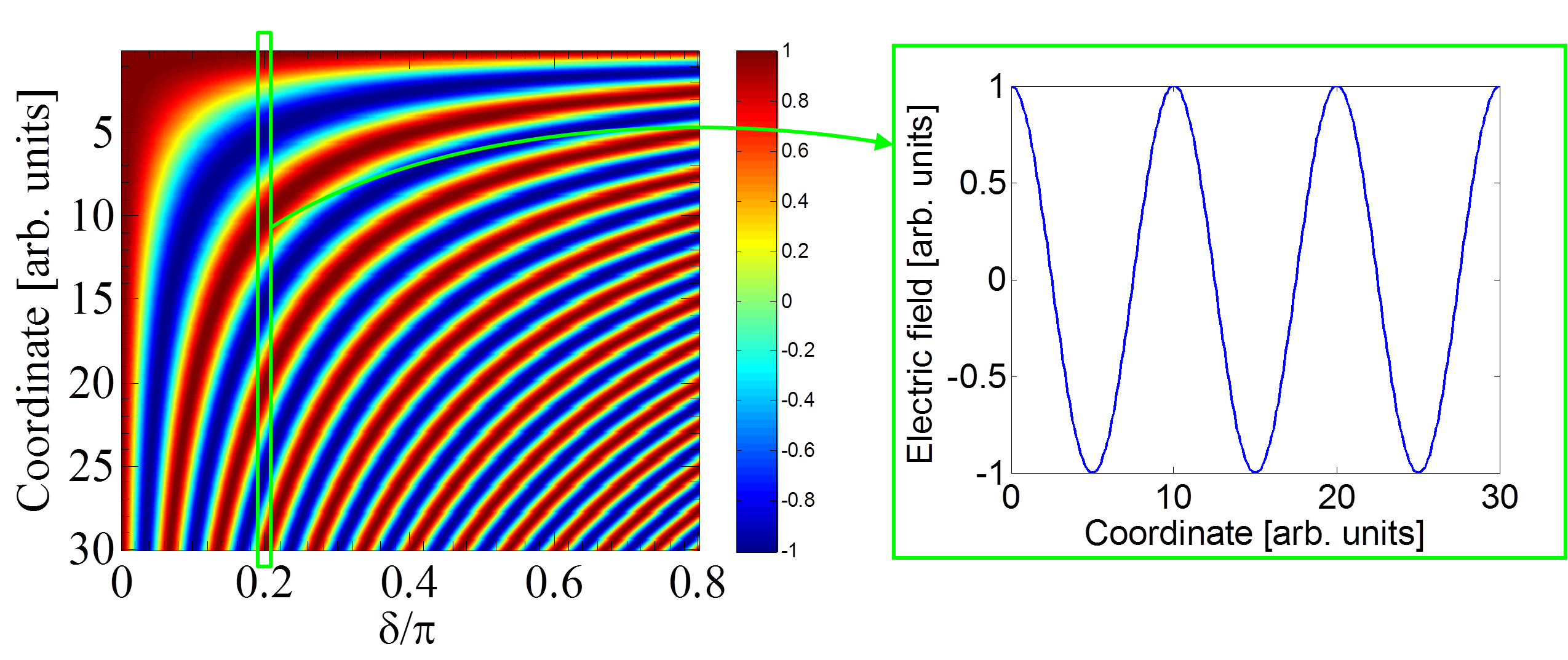}}
\caption{Left panel : the electric field amplitude $\omega-z$ map for free space. Right panel: a cross section of the map along the entire stack at a single frequency ($L/\lambda=3$).}
\label{figEmapfreespace}
\end{figure}

In this illustration, the entire structure is composed of type A slabs. The transmission spectrum
and the normalized density of modes are shown in figure (\ref{figTRhofreespace}).
As expected, both quantities are equal to one and independent of $\omega$. The electric field amplitude map is displayed in figure (\ref{figEmapfreespace}). The map shows the fundamental sinusoidal oscillations of the traveling plane wave, as seen in the cross section of the map on the right part of the figure.

\subsection{Fabry-Perot structure}

\begin{figure}[H]
\includegraphics[scale=0.27]{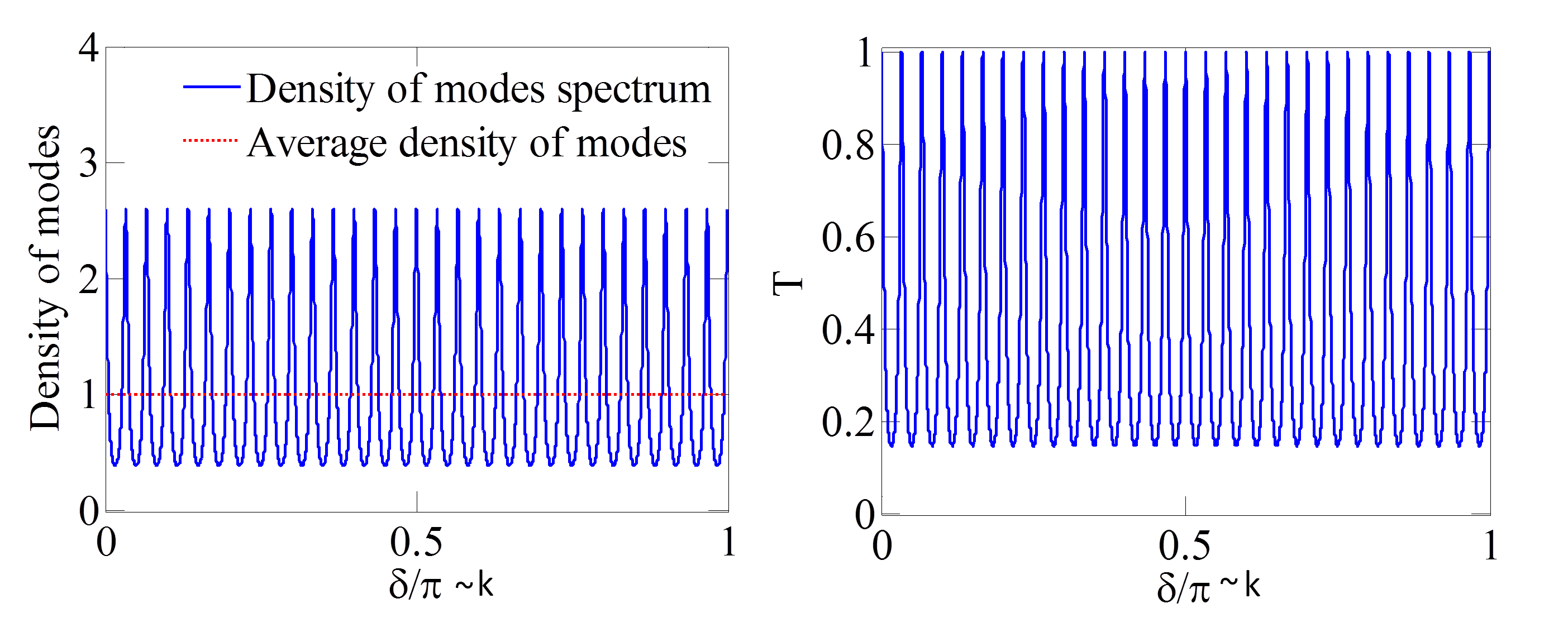}
\caption{The normalized density of modes (left part) and the transmission spectrum (right part), for a Fabry-Perot resonator.}
\label{figTRhoFP}
\end{figure}

\begin{figure}[H]
\includegraphics[scale=0.29]{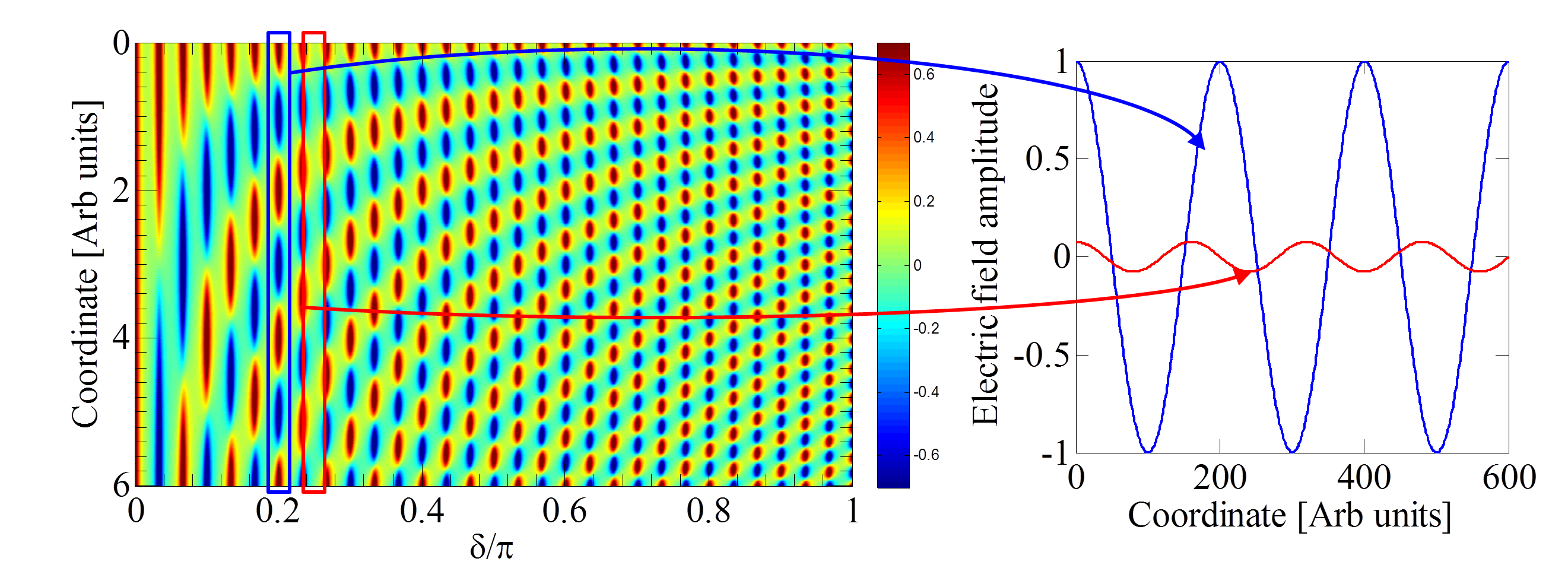}
\caption{Left panel : the electric field amplitude map for a Fabry-Perot resonator. Right panel:
a cross-section of the map along the entire stack at a frequency of
an allowed mode ($2L=6\lambda$ ) in blue and of a forbidden mode
($2L=7.5\lambda$ ) in red.}
\label{figEmapFP}
\end{figure}

The all dielectric Fabry-Perot resonator is realized here by placing two type B
slabs separated by several type A slabs, and for illustration we have chosen the refractive index $n_{B}$ to be three times larger than $n_{A}$. The transmission spectrum
and the normalized density of modes are presented in figure (\ref{figTRhoFP}).
Although this simple realization serves as a poor resonator,
transmission peaks and dips are distributed according to the Fabry-Perot resonance
condition $\frac{2L}{\lambda_{N}}=\left\{ n\right\} _{n=0,1,2...}$, where the density of modes is enhanced above the free one at resonances and attenuated off-resonance. This is, of course, due to interference effects inside the resonator as displayed by the corresponding electric amplitude map in  figure (\ref{figEmapFP}).

The electric field amplitude map is identical to the free space map with
the additional {}``selection rules'' for the Fabry-Perot cavity
("allowed" and "forbidden" modes). The two different cross sections of the map shown in the right part of figure (\ref{figEmapFP}) make clear the strong destructive interference effects taking place in 'off resonance' frequencies inside the structure.

\subsection{Periodic structure - photonic crystal}

%\subsubsection{Transfer matrix approach to periodic layered system}
\label{secperiodic}

The transfer matrix formalism introduced in section (\ref{sec:tmatrix}) is particularly convenient to study the case of a periodic layered system \cite{quasi,kohomotoothers,esaki}. In this case, the calculation of the total transfer matrix $\left(M_{cell}(\omega)\right)^N$ for a sequence of $N$ slabs, simplifies considerably thanks to the Cayley-Hamilton theorem which states that for unimodular matrices,
\be
M^2 - \left( \mbox{Tr} M \right) M + 1 =0
\label{ch}
\ee
This allows to easily obtain in  a recursive way, powers $M^n$ of the transfer matrix
\be
M^n (k) = U_{n-1} (\cos \theta) M(k) - U_{n-2} ( \cos \theta)
\label{MN}
\ee
where $U_n (\cos \theta) = {\sin (n+1 ) \theta \over \sin \theta}$ are the Chebychev polynomials of the second kind \cite{3terms}. A direct calculation shows that
\be
\mbox{Tr} M^n = \cos n \theta (k) = T_n (\cos \theta)
\ee
where $T_n (\cos \theta)$ are the Chebychev polynomials of the first kind. It is also straightforward to obtain the expression of the transmission coefficient $T_N = |t_N |^2$ for a Bragg system made of $N$ identical layers each described by the same transfer matrix $M$. From (\ref{MN}), we have the expression of $M^N$, which by definition is of the form (\ref{tmatrixsym}), namely,
\begin{eqnarray}
{1 \over t_N} &=& {U_{N-1} \over t} - U_{N-2} \nonumber \\
{r_N  \over t_N} &=& {r \over t} - U_{N-1}
\label{rec}
\end{eqnarray}
\begin{figure}[H]
\includegraphics[scale=0.26]{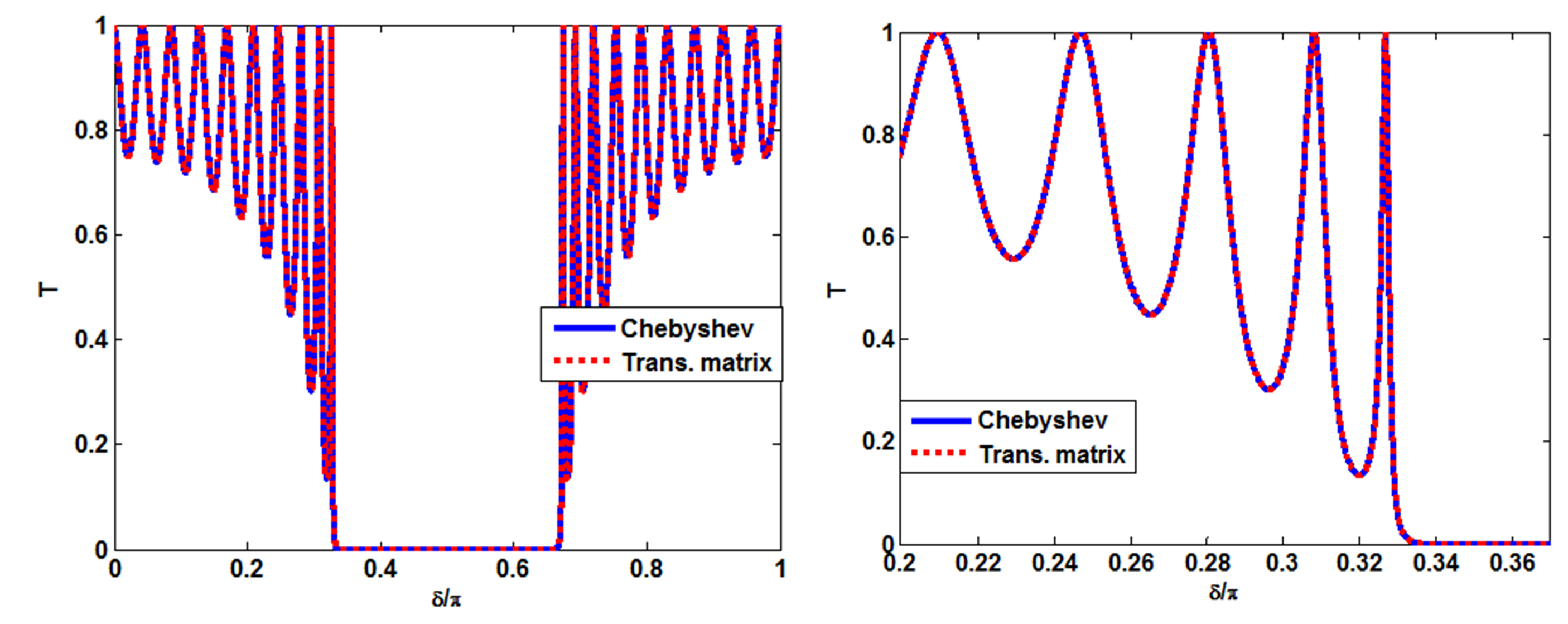}
\caption{Periodic Bragg system. The band structure is calculated using both the Chebychev polynomials (\ref{bragg1}) and via the transfer matrix approach (\ref{rhoofomega}). Right part - a close up on the long wavelength band edge mode.}
\label{figPCchevyTmatrix}
\end{figure}
Since $\mbox{det} \, M^N = 1$, then $\left| {1 \over t_N} \right|^2 - \left| {r_N  \over t_N}  \right|^2 = 1$, so that
\be
T_N = |t_N |^2 = {1 \over 1 + \left| {r  \over t}  \right|^2 \left( {\sin N \theta \over \sin \theta } \right)^2 }
\label{bragg1}
\ee
This is a straightforward way to numerically obtain the well known Bragg spectrum (see Figs. \ref{figPCchevyTmatrix} and \ref{figPCT}).
\begin{figure}[H]
\includegraphics[scale=0.35]{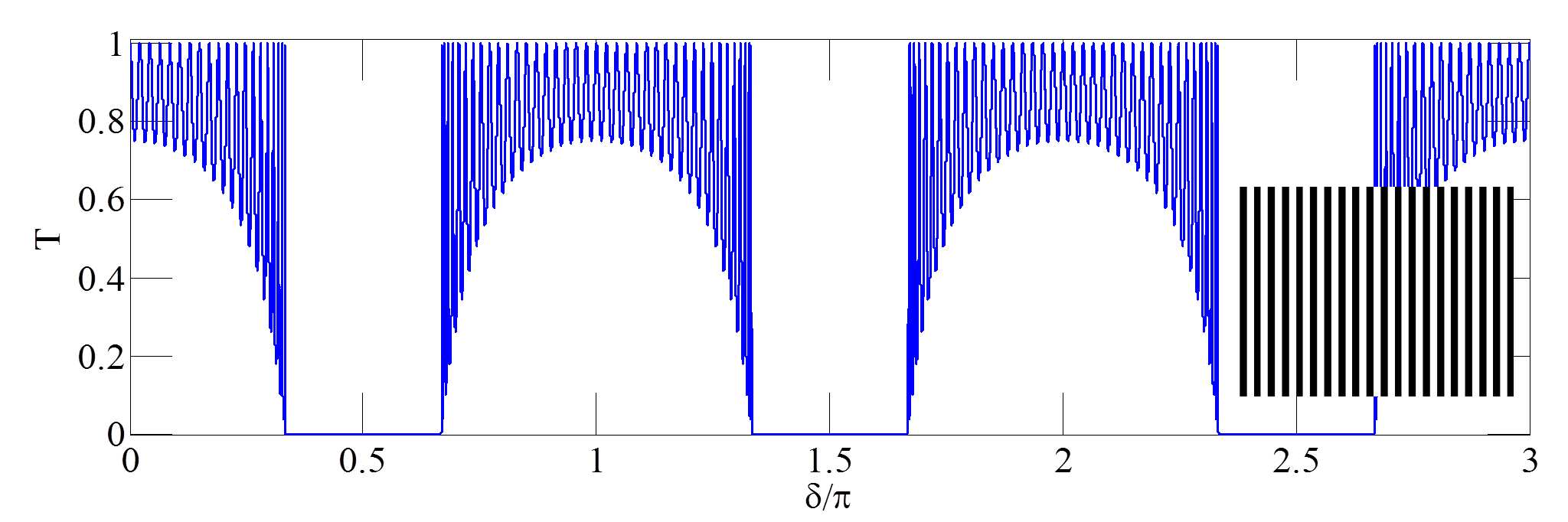}
\caption{The transmission spectrum for a 40 slab long photonic crystal whose structure is illustrated in the inset.}
\label{figPCT}
\end{figure}
\begin{figure}[htb]
\includegraphics[scale=0.35]{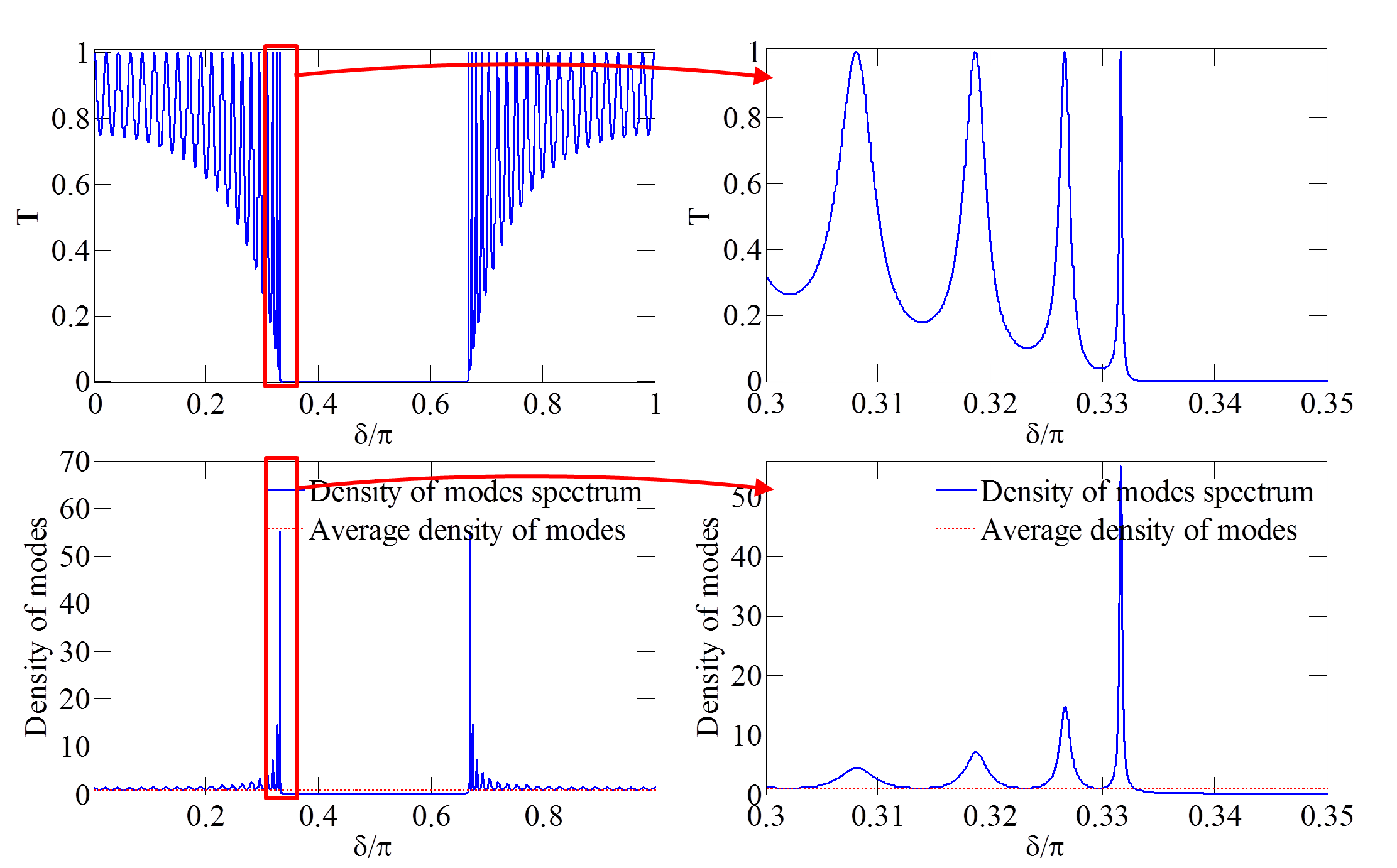}
\caption{The transmission spectrum (top part) and the normalized density of modes (bottom part) for a 40 slab photonic
crystal. The figures on the right show a close-up on the long wavelength band edge mode}
\label{figPCTRho}
\end{figure}

\medskip

\section*{Remark}
The transmission coefficient $T_N(\theta)$ for a finite periodic system (10 unit cells in length) were calculated using both a direct transfer matrix calculation and the relation (\ref{theequation}) and more directly (\ref{bragg1}). Both are plotted in figure (\ref{figPCchevyTmatrix}) and the agreement is perfect. From the relation (\ref{bragg1}), it appears clearly that the resonances are given by the condition ${\sin N \theta \over \sin \theta } = 0 $. Note that this is specific to the periodic  case and cannot be generalized to other more complex structures, for which the general relation (\ref{theequation}) must be used to obtain the spectrum.
\medskip

%\subsubsection{Transfer matrix approach to periodic layered system}

The dielectric photonic crystal analyzed from here forth is composed of alternating type A and
B slabs with 20 repetitions of the unit cell [AB]. The refractive index is chosen to be $n_{B} = {3 \over 2} n_A$.
 The transmission spectrum for this 40 slab photonic crystal
is shown in Figure \ref{figPCT}
The periodicity gives rise to  transmission bands and photonic band
gaps centered at the $\frac{\lambda}{4}$ condition, as expected. A
closer look at the transmission and the density of modes spectrum near
the first stop band is depicted in Figure \ref{figPCTRho}

The sharp band edge modes with a 50-fold enhancement in the DOM compared to a free system can
be seen in the close up figures. The number of resonant modes corresponds
to the number of unit cells.
%\medskip

{\bf Remark :}
All spectral quantities presented here were calculated using the total transfer or the total scattering matrix, but are are completely equivalent to (\ref{bragg1}).
%\medskip

\begin{figure}[htb]
\includegraphics[scale=0.27]{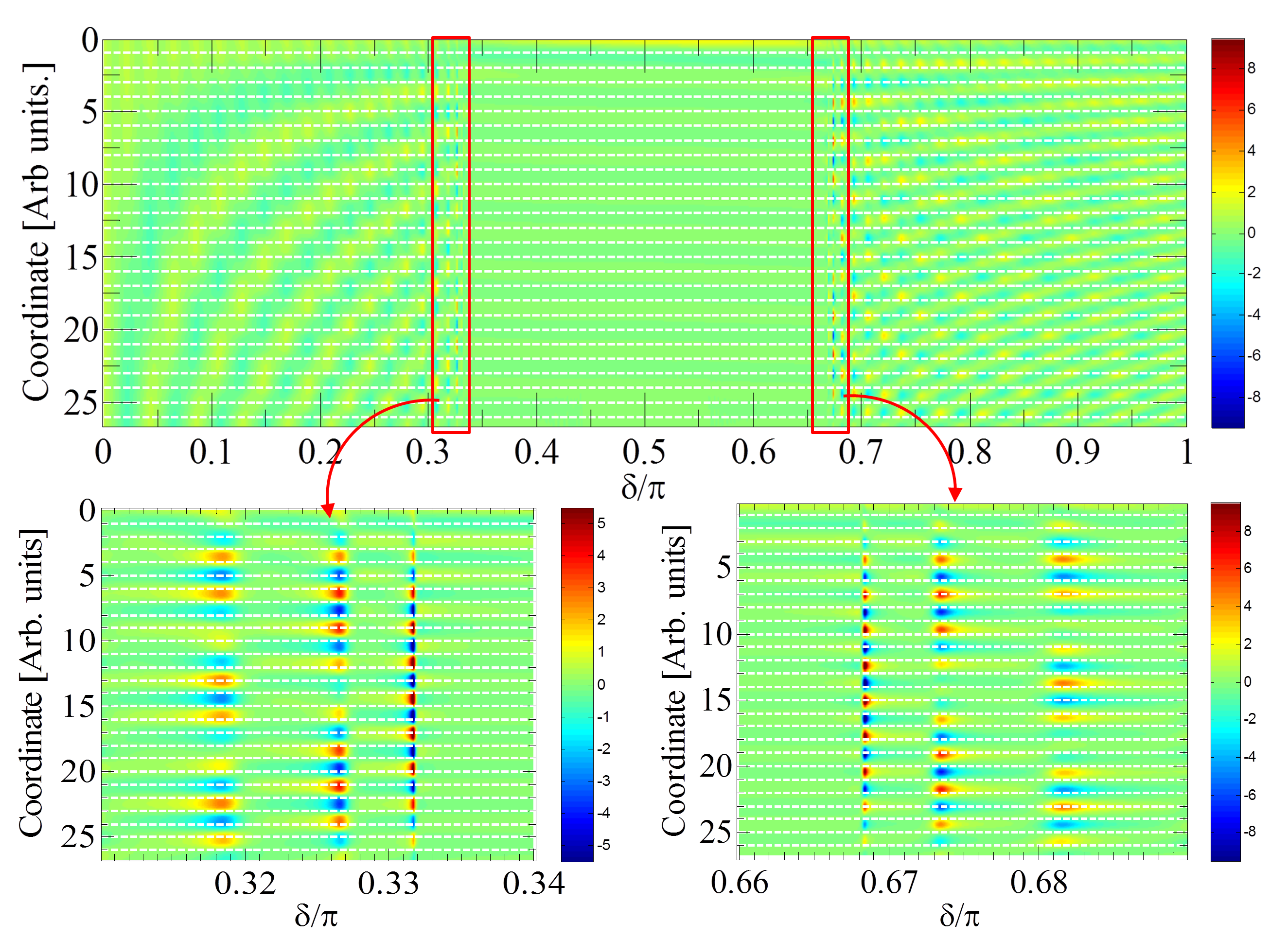}
\caption{The electric field amplitude $\omega-z$ map for a 40 slab photonic crystal. Spatial boundaries of each slab  are indicated by dashed gray lines. Bottom figures are a close-up on the long (bottom left figure) and short (bottom right figure) band edge modes.}
\label{figPCEmap}
\end{figure}

\begin{figure}[htb]
\includegraphics[scale=0.26]{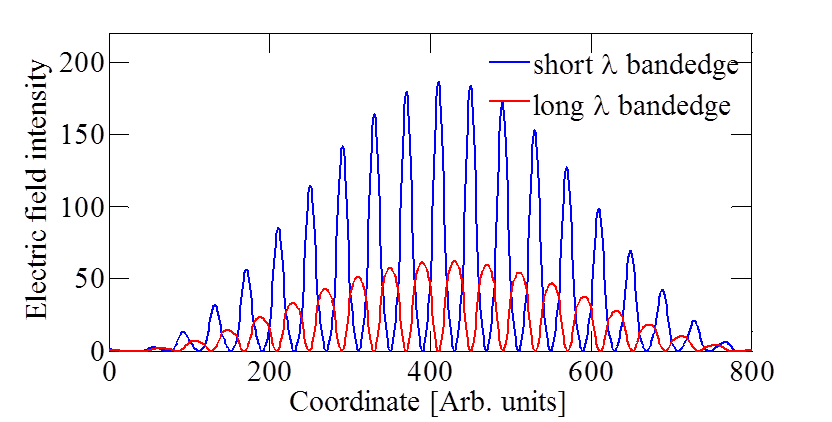}
\caption{the electric field amplitude cross section for the long and short wavelength band edge modes in a 40 slabs photonic crystal. Electric field of the short and long wavelength modes concentrate on different slab types.}
\label{figPCECS}
\end{figure}

The electric field
amplitude map corresponding to this photonic crystal is displayed in Figure \ref{figPCEmap}, as well as a close up look at the
band edge modes.
Figure \ref{figPCEmap} shows that the band edge modes also have an enhanced electric
field distribution peaks - up to 12 times that of the incoming field, due
to constructive interference effects and transient energy buildup. It is also visible that while the electric field "hot-spots"
in the long wavelength band edge mode tends to be located in one type
of slab, it tends to be located
in the other type for the short wavelength band edge mode. This feature is presented more clearly in Figure \ref{figPCECS}.

\subsection{Random structure}

Disordered layered structures have been studied in very great details as a successful and easy to implement model for 1D realization of Anderson localization \cite{erdos,gredeskul,luck} (and references therein) and other related complex systems. It would be a hopeless task to list all relevant works here and it is also not our purpose to study this class of systems in greater details. We discuss them in this review mostly  for the sake  of completeness and also to present results for the electric field amplitude map not so often discussed in the existing literature.

\begin{figure}[htb]
\includegraphics[scale=0.28]{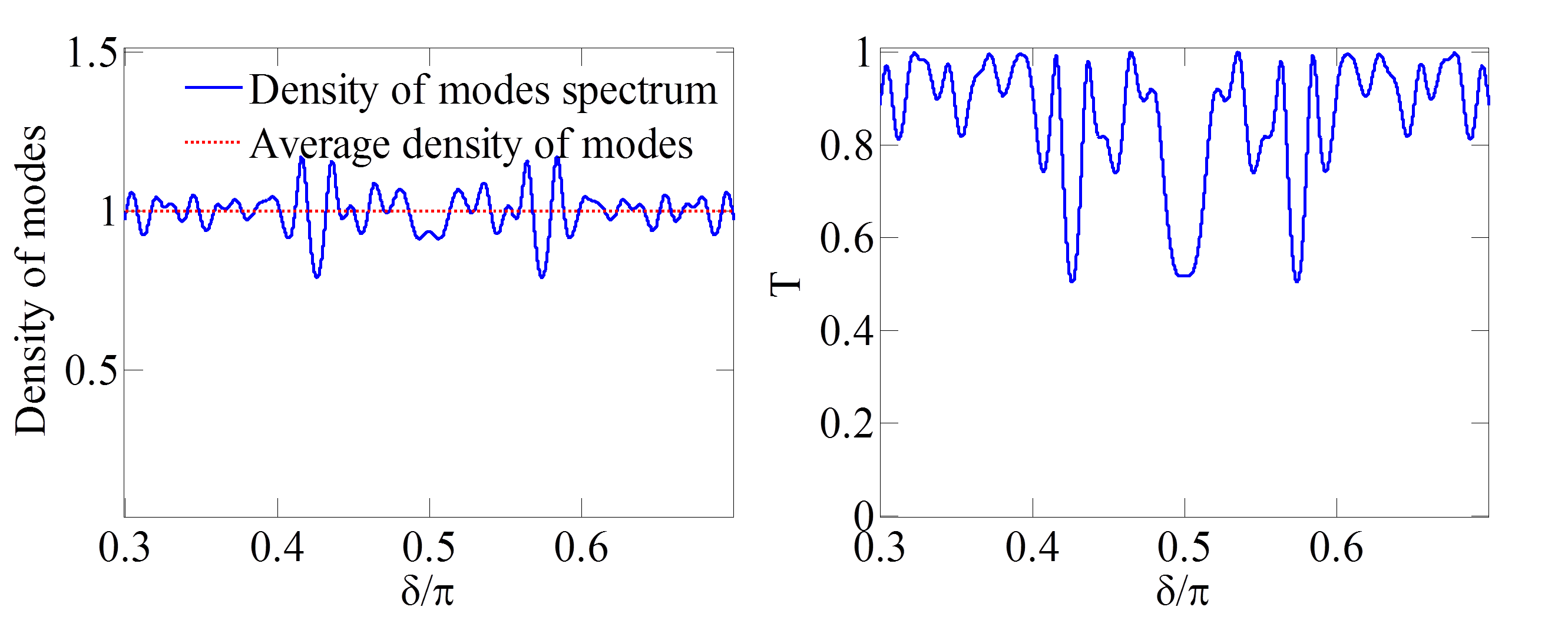}
\caption{The normalized density of modes (left part) and the transmission spectrum (right part), for a weakly disordered structure}
\label{figDOweakTRho}
\end{figure}

\begin{figure}[H]
\includegraphics[scale=0.28]{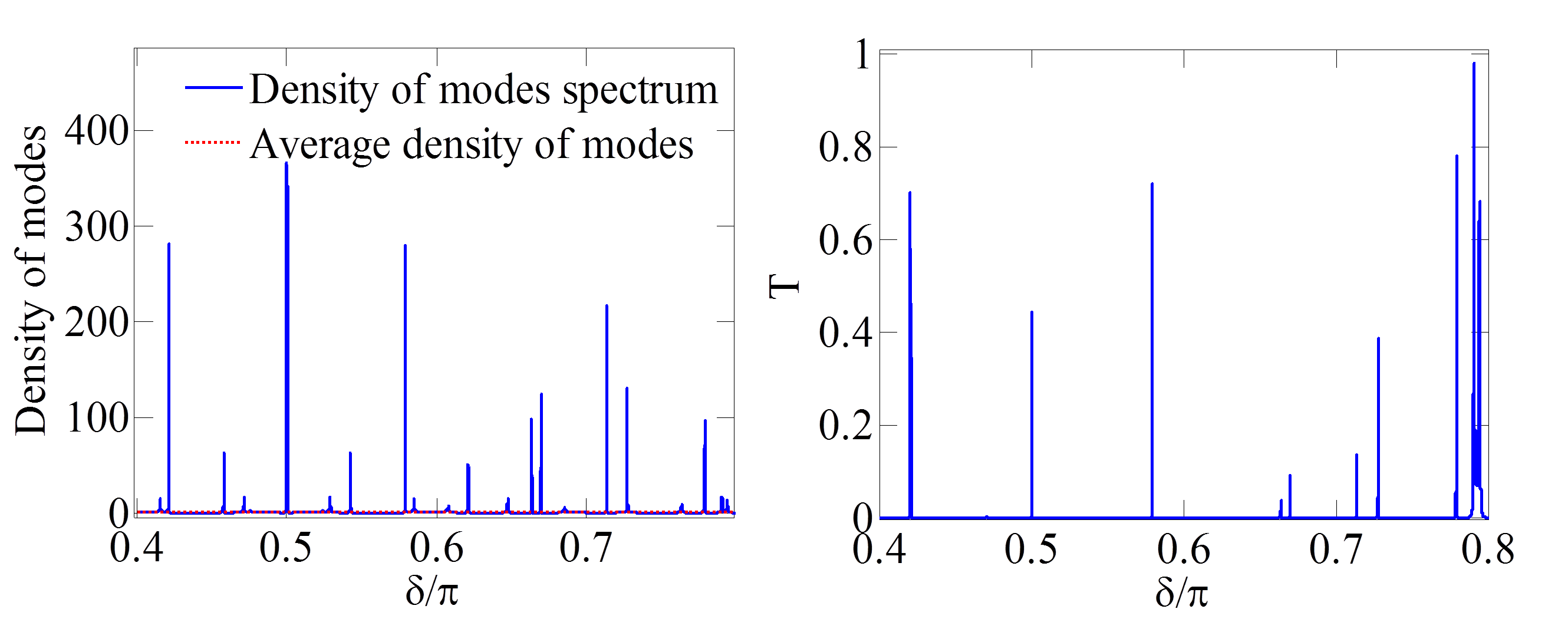}
\caption{The normalized density of modes (left part) and the transmission spectrum (right part), for a strongly disordered
structure}
\label{figDOstrongTRho}
\end{figure}

A dielectric random system is generated here by a random sequence of type A and B slabs. Weak disorder is obtained by setting the refractive index $n_{B}$ to be 10\% larger than $n_{A}$, and for strong disorder, the refractive index contrast is set as $n_{B} = 2 n_{A}$.
The transmission spectrum and the normalized density of modes for
a weakly disordered structure is displayed in figure \ref{figDOweakTRho}.
It can be seen in the right part of figure \ref{figDOweakTRho} that a weak disorder can attenuate the
transmission spectrum significantly, but as seen in the left part of figure \ref{figDOweakTRho}, the density of modes
shows no localized modes - the localization length in this case
is greater than the total length of the structure, so the modes appear {}``smeared''
into each other. The transmission spectrum and the normalized density
of modes for strong disorder is shown in figure \ref{figDOstrongTRho}

\begin{figure}[ht]
\includegraphics[scale=0.32]{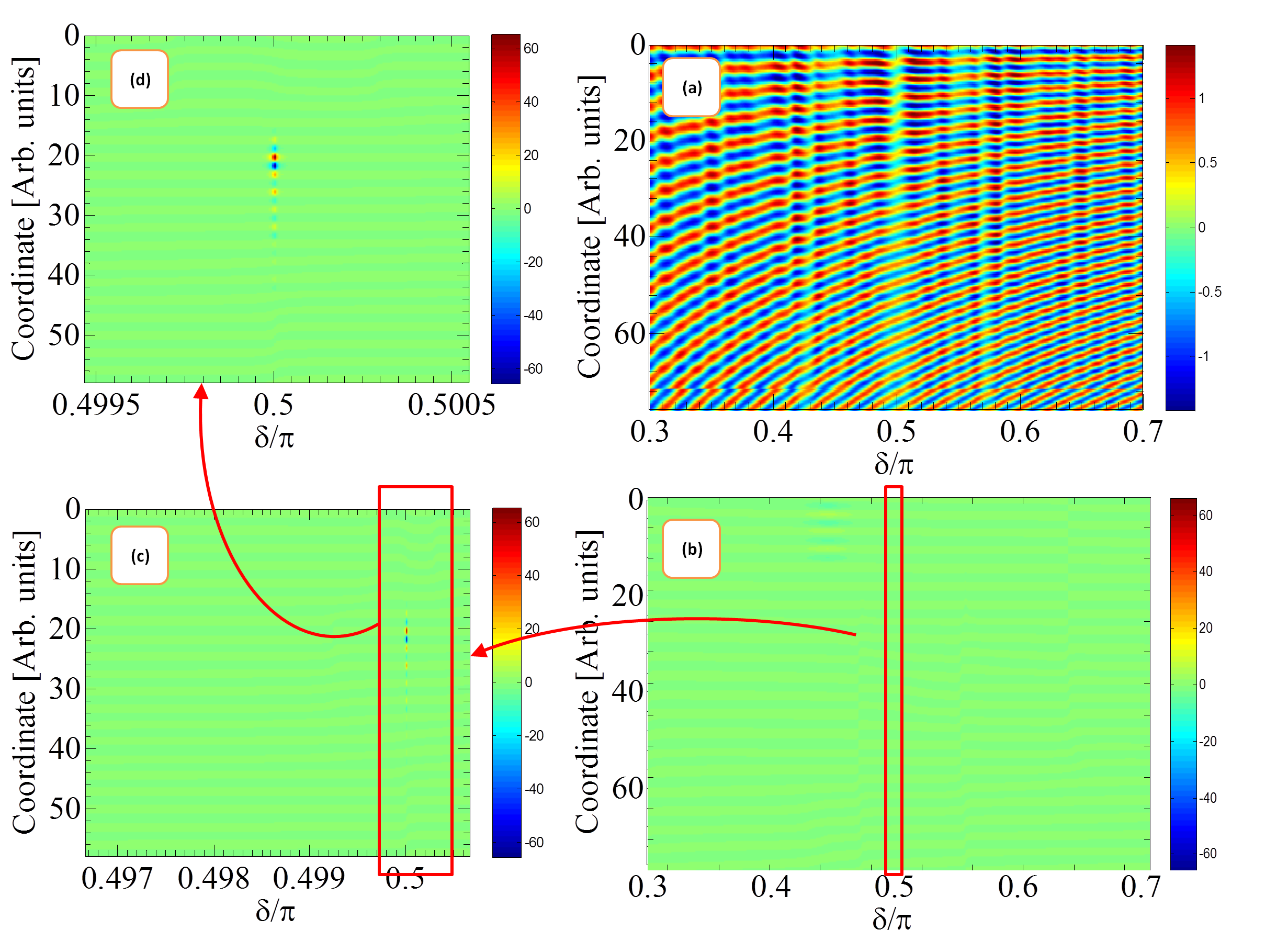}
\caption{The electric amplitude map for disordered structure. a: weakly disordered structure.
b-d: strongly disordered structure (c-d are zooming on a resonant mode).}
\label{figDOEmap}
\end{figure}

In contrast to the previous case of weak disorder, the transmission for strong disorder
is inhibited for all frequencies except for a set of discretely distributed modes.
The density of modes shows the same discrete resonant frequency pattern
together with a significant enhancement of the DOM campared to a free system. The electric field amplitude
map for weakly and strongly disordered structures is shown in figure \ref{figDOEmap}.

The modes of a weakly disordered structure shown in subplot (a) of figure \ref{figDOEmap} are smeared spatially and the spatial distribution of electric field extends
to the edges of the stack. For the case of strong disorder in subplot (b) of figure \ref{figDOEmap}, the electric field amplitude
map displays vanishingly small values for all frequencies except some discrete isolated
modes as the one shown in subplot (d) of figure \ref{figDOEmap}. These modes are spatially localized with a localization length
smaller then the structure length. The electric field amplitude is
greatly enhanced in the resonant modes - about 70 times that of the
incoming wave - also due to constructive interference and transient energy buildup.

\subsection{Fibonacci structure}

Among the large variety of complex structures, quasi-periodic distributions play a special role. One of the many reasons for this ongoing and continuous interest, is that quasi-periodic structures are deterministic (non random)  but nevertheless display interesting qualitative similarities (but quantitative distinct features) with random structures, {\it e.g.}  localization of the modes, insulating behavior, structured spectra\cite{marcia,damanik}, $\dots$.

Moreover, quasi-periodic 1D structures can be viewed as realizations of quasi-crystalline  order and there are interesting candidates to study transport and thus understand the insulating behavior of these structures. In this short review, we consider a two-word Fibonacci and a Cantor set structure as  working examples to test our approach to obtain spectral properties using the $S$-matrix as in (\ref{theequation}). An interesting point about the Fibonacci case is that while this real space structure is not fractal (it does not display a discrete scaling symmetry leading to a self-similar order), the corresponding spectral quantities display a self-similarity property yet to be fully defined \cite{luck,quasi,damanik,damanik2,kohomotoothers}.

The literature on quasi-periodic distributions is extremely rich. Here we present some references which may be considered as leads to a more complete bibliography. The initial interest in such problems traces back to works by Dyson and Schmidt \cite{luck}. For a recent review on spectral properties see \cite{marcia}, and \cite{damanik}. Applications of quasi-periodic structures to electronic transport properties and to electromagnetic wave propagation are numerous. More closely related to the topics covered in this book are \cite{marcia, quasi, dalnegro1, dalnegro2,kohomotoothers}.

\begin{figure}[ht]
\includegraphics[scale=0.32]{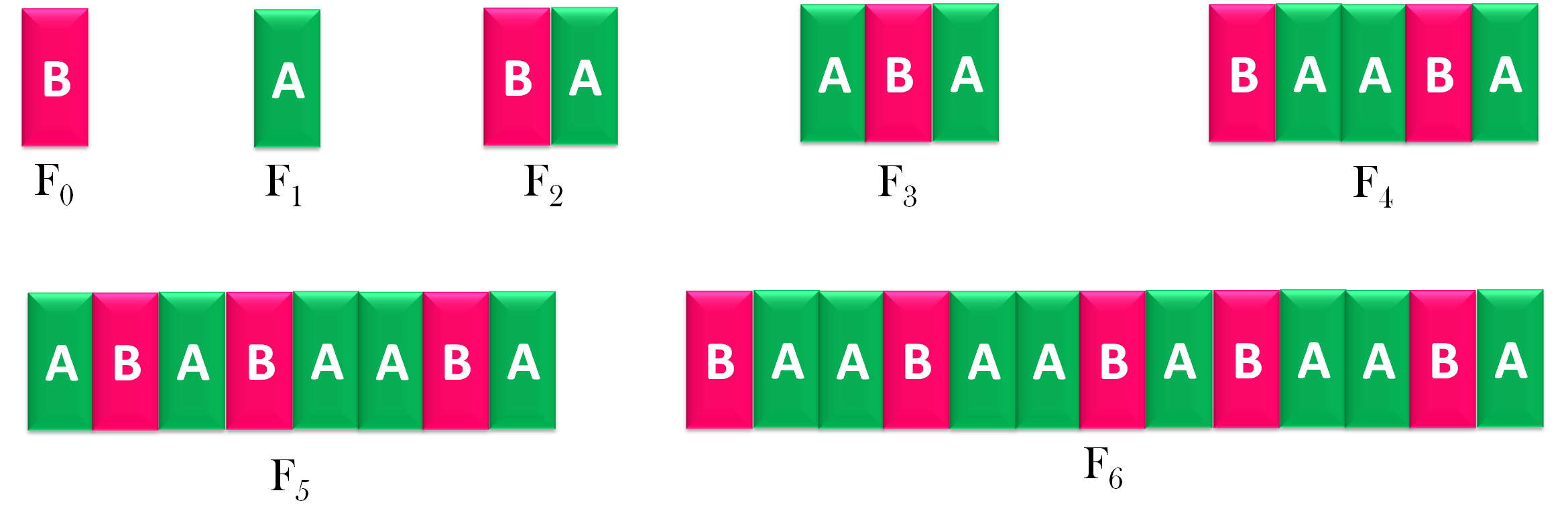}
\caption{A schematic view of generations 0-6 of the Fibonacci structure used in the calculations}
\label{figFiboseq}
\end{figure}

\begin{figure}[ht]
\includegraphics[scale=0.28]{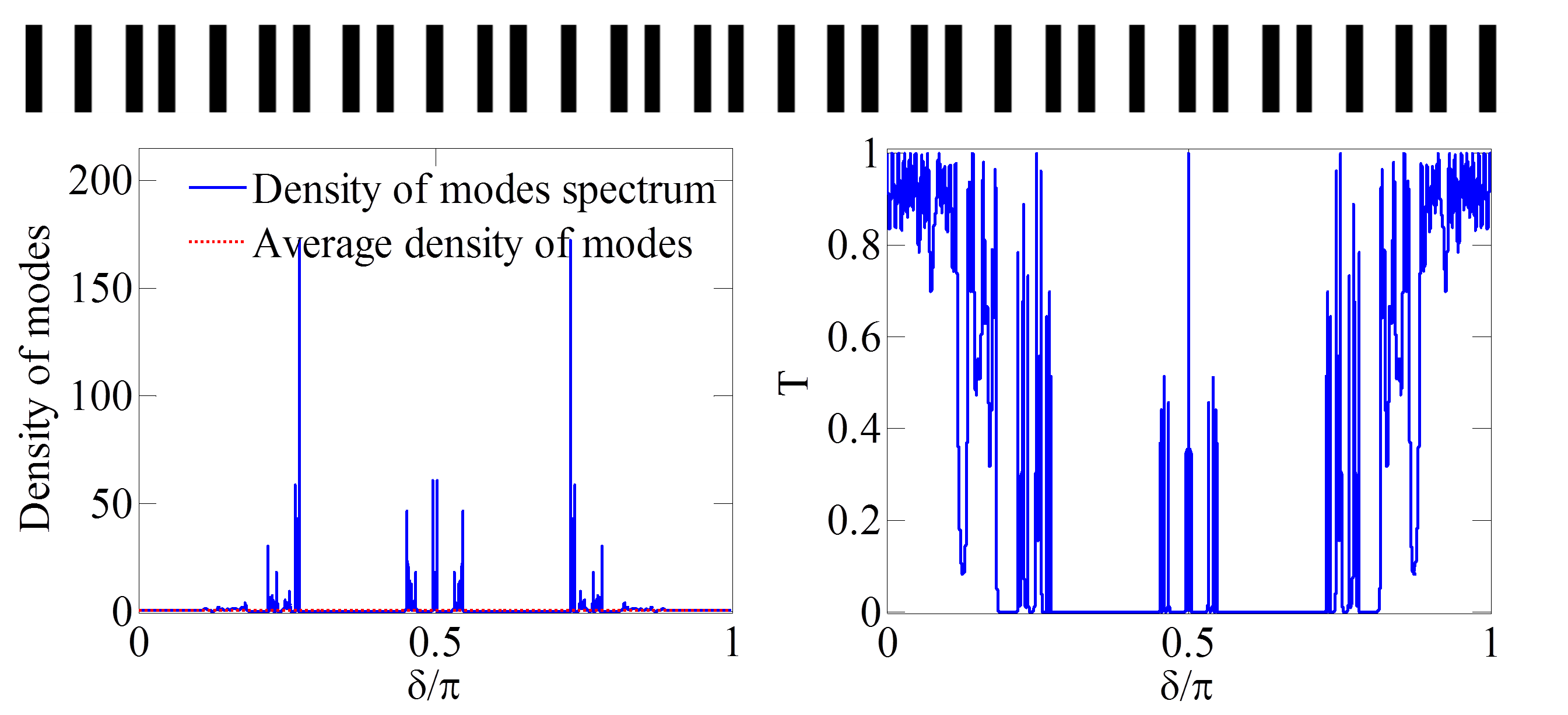}
\caption{The normalized density of modes (left part) and the transmission spectrum (right part), for the 10th generation Fibonacci structure.
The structure of the stack is illustrated above the two spectra.}
\label{figFiboTRho}
\end{figure}

The Fibonacci structure can be defined in many different ways. Generally, consider an alphabet $\mathcal{A} = \{ A,B,C, \cdots\}$ composed of different letters, and a substitution rule $\xi \left( \mathcal{A} \right)$, so that
\be
\xi \left( ABC \cdots \right) = \xi (A) \, \xi(B) \, \xi(C)\, \cdots
\label{sub}
\ee
Then, a Fibonacci sequence is $F_n = \xi^n (W_0 )$, where $w_0$ is some initial word. Her, we consider an alphabet composed of two letters $A$ and $B$ namely slabs of type A and type B arranged according to a Fibonacci sequence $\xi (A) = BA$, $\xi(B) = A$ and $F_n = \xi^n (B)$, namely,
\be
A \rightarrow BA  \rightarrow ABA \rightarrow BAABA \rightarrow ABABAABA \rightarrow \cdots
\label{fibex}
\ee
Alternatively, the same structure can be defined by the first two words, and a building rule stating that the third word and on is composed of the sum of the previous two words, namely,
%Among the many possibilities to generate a Fibonacci sequence {\color{blue}ref??}, we consider here the following Fibonacci building rule:
\be
F_{0}=B;\; F_{1}=A;\; F_{j>1}=[F_{j-2}F_{j-1}]
\label{ch}
\ee
An illustration of the resultant structures is depicted in figure \ref{figFiboseq}. Setting the refractive
indices to be $n_{B} = {3 \over 2} n_{A}$, the transmission spectrum and the normalized density of modes for 10th
generation Fibonacci structure are shown in figure \ref{figFiboTRho}.

\begin{figure}[H]
\includegraphics[scale=0.28]{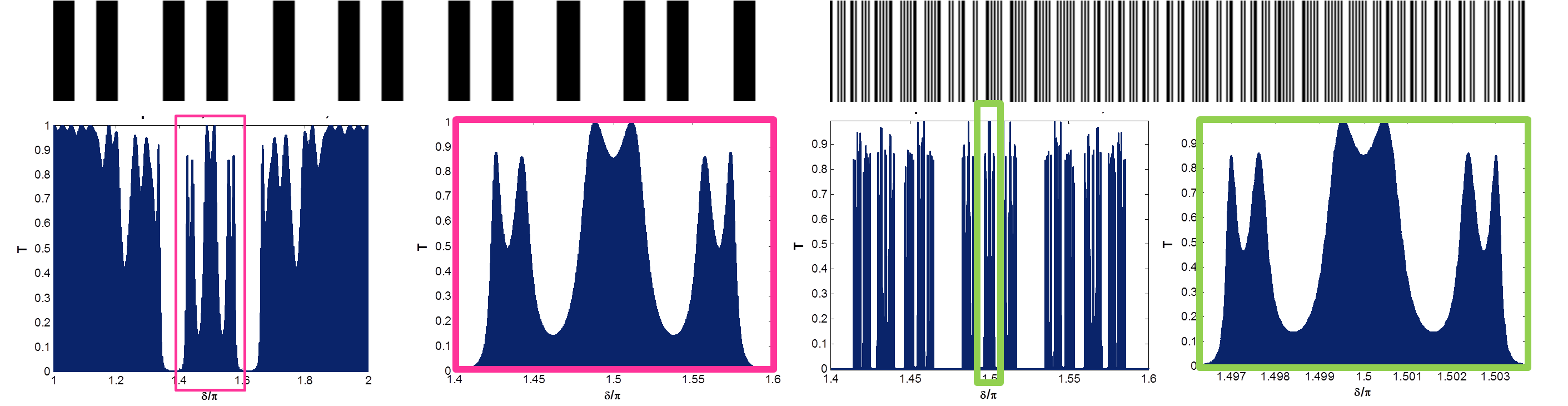}
\caption{Transmission spectrum and stack structure for 8th generation (left part),
and 14th generation Fibonacci structure (right part). Structures are illustrated above the spectra. The self similarity of the spectra is evident in the close-up on the middle part of the two spectra.}
\label{figFiboss}
\end{figure}

\begin{figure}[H]
\includegraphics[scale=0.32]{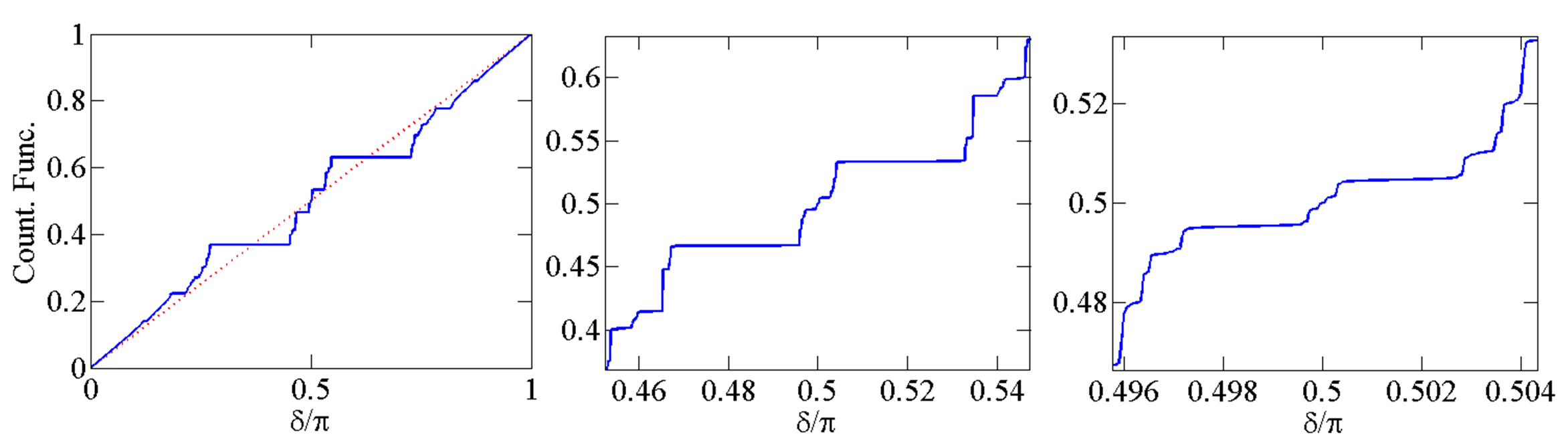}
\caption{the counting function $\mathcal{N}\left(\omega\right)$ for a 10th generation Fibonacci structure. The middle and right figures are a gradual zoom on the counting function depicted on the left figure. A qualitative self similarity is observed.}
\label{figFiboCF}
\end{figure}

The spectrum shows two distinct band gaps, and some spectrally isolated
band edge modes with enhanced DOM, arranged according to a self-similar ``fractal like'' structure. Note that the DOM does not obey an exact self-similarity, but only an approximate one. However, it has been established \cite{damanik,damanik2} that the energy spectrum of an infinitely long Fibonacci structure is multi-fractal from the cantor set family (see figure \ref{figCantorseq}).
To explore this self similarity, we consider the transmission spectrum
for two different generations of the structure. The spectrum is depicted
in figures \ref{figFiboss} and \ref{figFiboCF}.

\begin{figure}[H]
\includegraphics[scale=0.36]{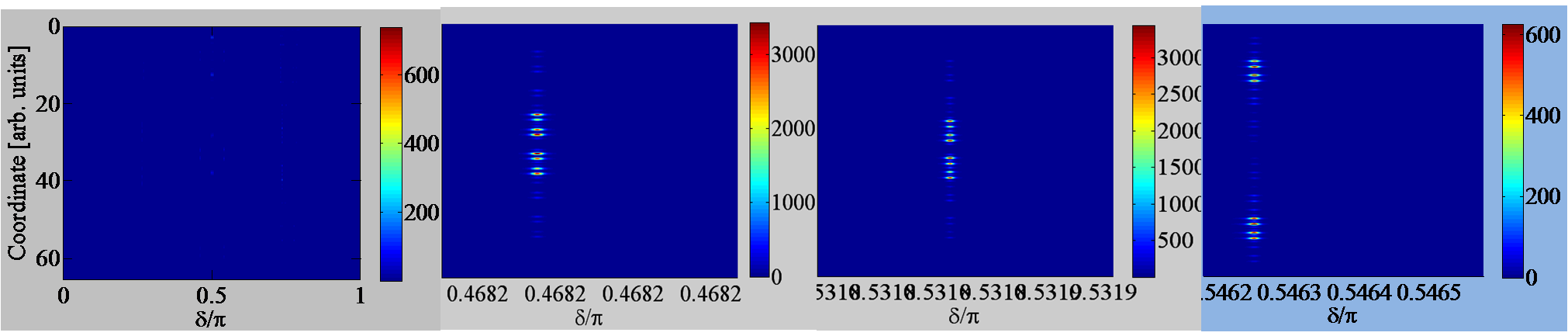}
\caption{The electric field intensity map for the 8th generation Fibonacci structure.
Left- full spectrum. Other plots: closeup zoom on some isolated modes. Spatial localization and spatial self-similarity are apparent.}
\label{figFiboEmap}
\end{figure}

\begin{figure}[H]
\includegraphics[scale=0.37]{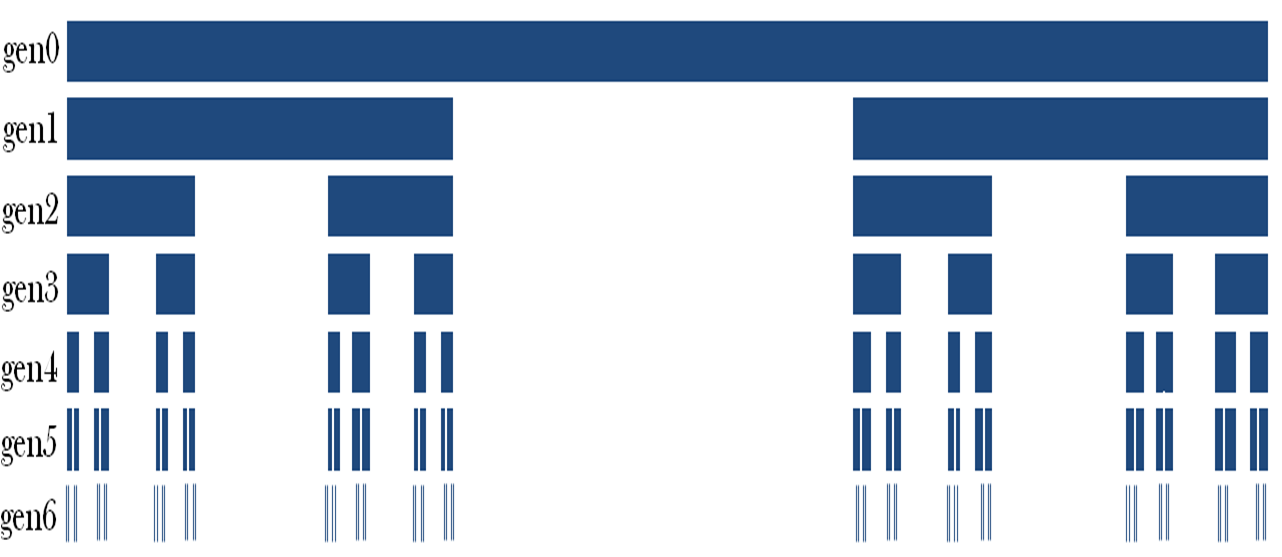}
\caption{A schematic view of generations 0-6 of a triadic cantor set.}
\label{figCantorseq}
\end{figure}

The similarity between the middle section of the transmission spectra
for different Fibonacci structure can be seen in the close up figures
(the middle spectral feature is the same for very different bandwidths).
This result initially obtained in \cite{quasi,kohomotoothers} is reproduced here.
The electric intensity map for the 8th generation Fibonacci structure,
and a close up look at some interesting modes is shown in
figure \ref{figFiboEmap}.
The plots in figure \ref{figFiboEmap} show that the electric field intensity is vanishingly small for
almost all frequencies except for some discrete modes. For some of these modes
the electric intensity spatial distribution is localized and (surprisingly)
symmetric around the middle of the stack, and self similar as it resembles
the Cantor set structure - see figure \ref{figCantorseq}. The reason for the symmetry is believed to
be the palindromic nature of the Fibonacci sequence \cite{damanik2}. This somewhat odd electric field spatial structure is probed due to the enhanced intra-slab resolution of the calculation. Here too there is a strong enhancement of the electric field amplitude in the band edge modes - up to 60 times that of the incoming field.

\subsection{Fractal Cantor set structure}

\begin{figure}[H]
\includegraphics[scale=0.28]{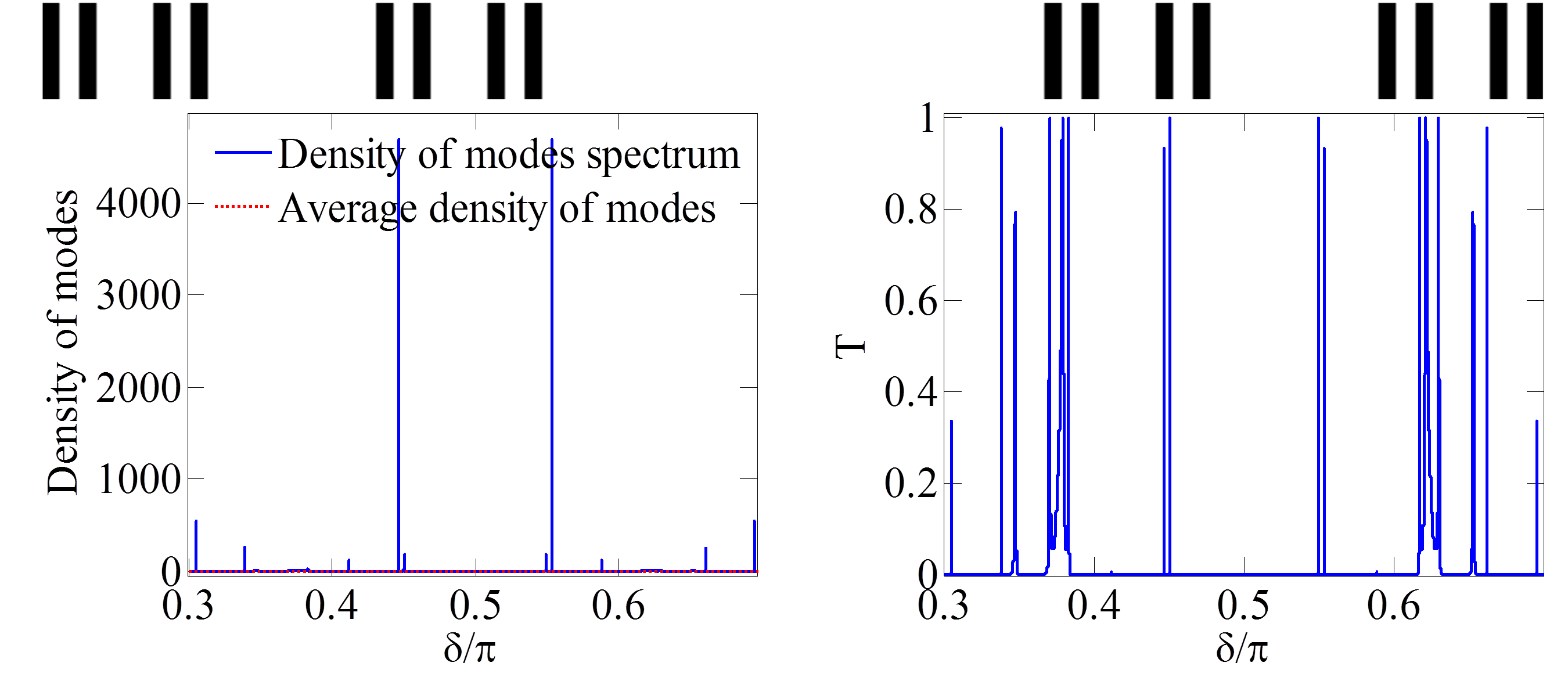}
\caption{The normalized DOM (left part) and the transmission spectrum (right part), for the 4th generation triadic Cantor set structure. The structure of the stack is illustrated above the two spectra.}
\label{figCantorTRho}
\end{figure}

\begin{figure}[H]
\includegraphics[scale=0.28]{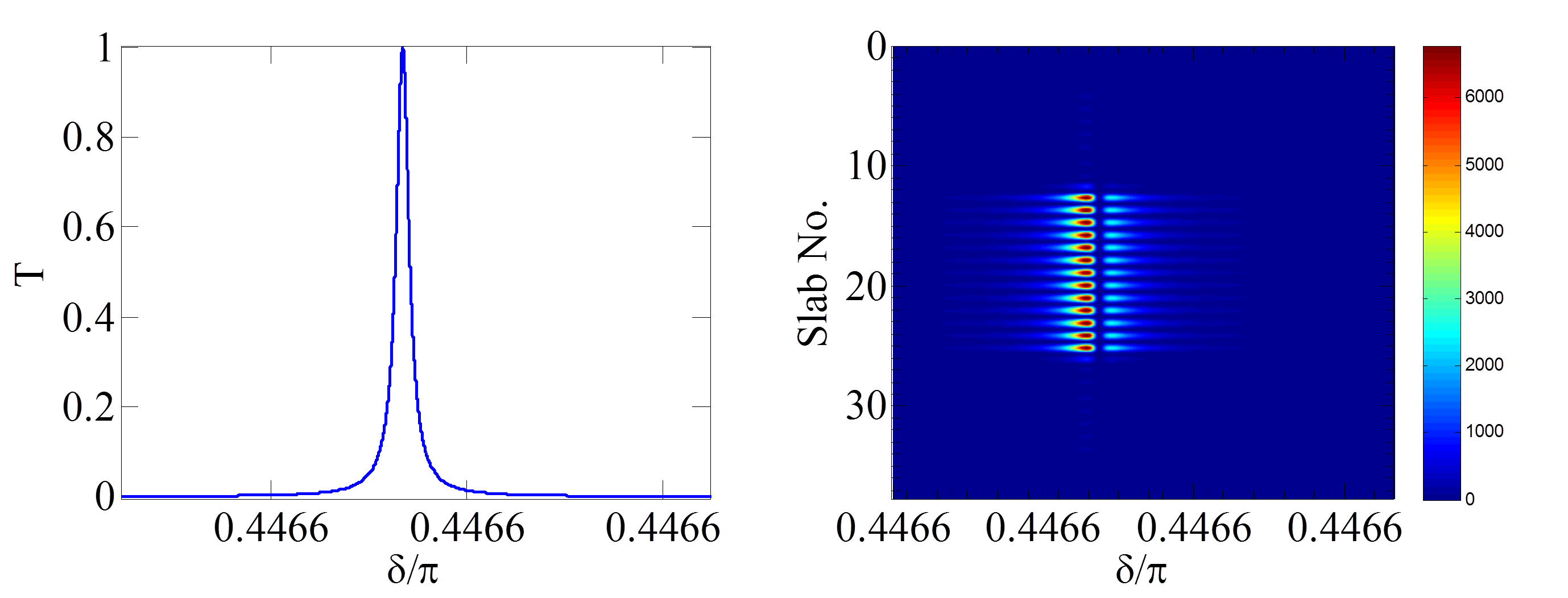}
\caption{A close-up on the transmission curve (left) and electric field intensity map (right) for a resonant mode in a 4th generation Cantor set structure.}
\label{figCantorEmap}
\end{figure}

The Cantor set structure is realized here by arranging type A and type
B slabs according to the triadic Cantor sequence (see figure \ref{figCantorseq}), and setting the
refractive index $n_{B}$ to be 50\% larger than $n_{A}$. The transmission
spectrum and the normalized density of modes for a 4th generation Cantor set structure
is shown in figure \ref{figCantorTRho}.
Figure \ref{figCantorTRho} shows that as for the random and the Fibonacci structures, this structure too possesses a jagged energy spectrum
with discrete resonant modes accompanied by a DOM enhancement. A close
up look at the transmission curve of one of these modes, and the electric
field intensity map for the same mode is given in figure \ref{figCantorEmap}.

Figure \ref{figCantorEmap} shows that as for the Fibonacci structure, the resonant
modes in a Cantor set structure are isolated spectrally, and localized
spatially. However, the spatial localization in the Cantor case is
geometric in nature - the electric field spatial distribution is enclosed
in the middle third of the structure, while the smaller Cantor motifs at both sides
serve as dielectric mirrors.

\subsection{Defect modes in photonic crystal structure:}

\begin{figure}[H]
\includegraphics[scale=0.28]{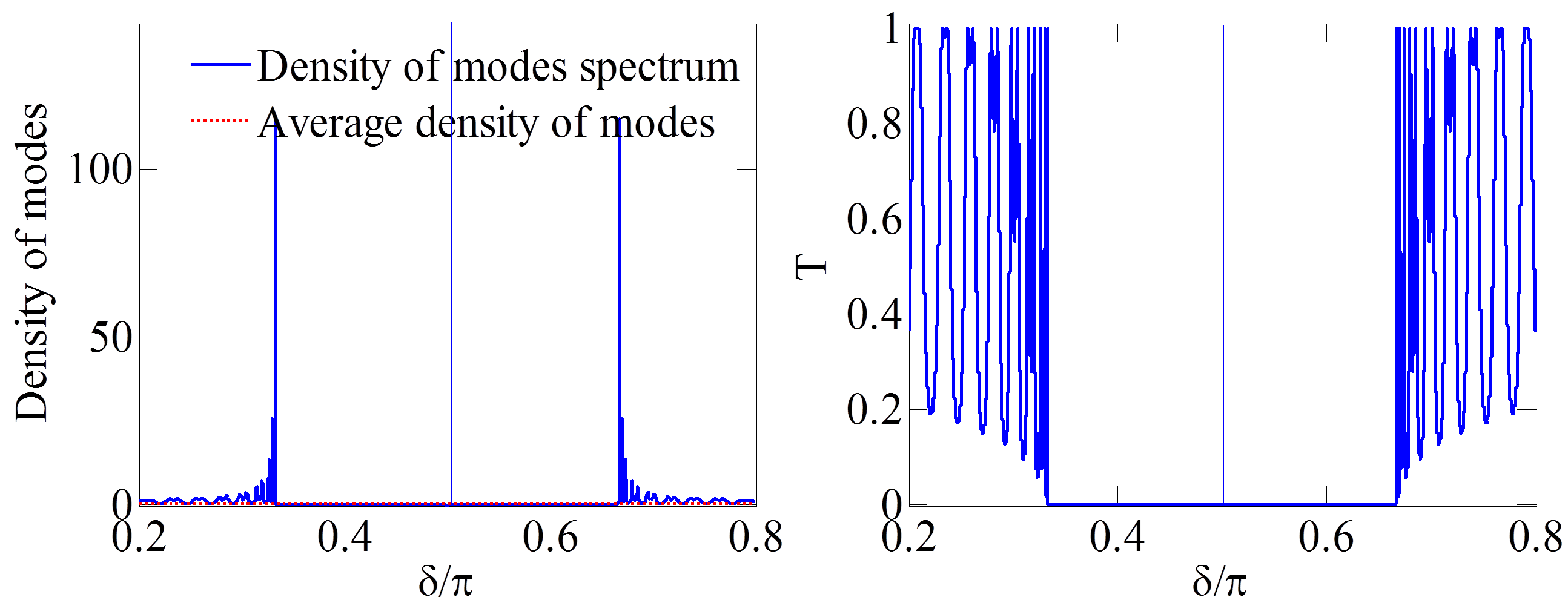}
\caption{The normalized density of modes (left part) and the transmission spectrum (right part), for a photonic crystal hosting a single defect.}
\label{figPCDMTRho}
\end{figure}

The Fibonacci and Cantor set structures are examples for quasi-periodic
structures which give rise to deterministic modes isolated spectrally and
localized spatially. The {}``work horse'' of this kind of modes
is the defect mode - a defect (say an [AA] segment) is inserted to the center of a photonic
crystal to produce a mid gap mode, localized spatially inside the
defect\cite{pbg}. The transmission spectrum and the normalized density of modes for a photonic crystal hosting a single impurity
is shown in figure \ref{figPCDMTRho}
\begin{figure}[H]
\includegraphics[scale=0.45]{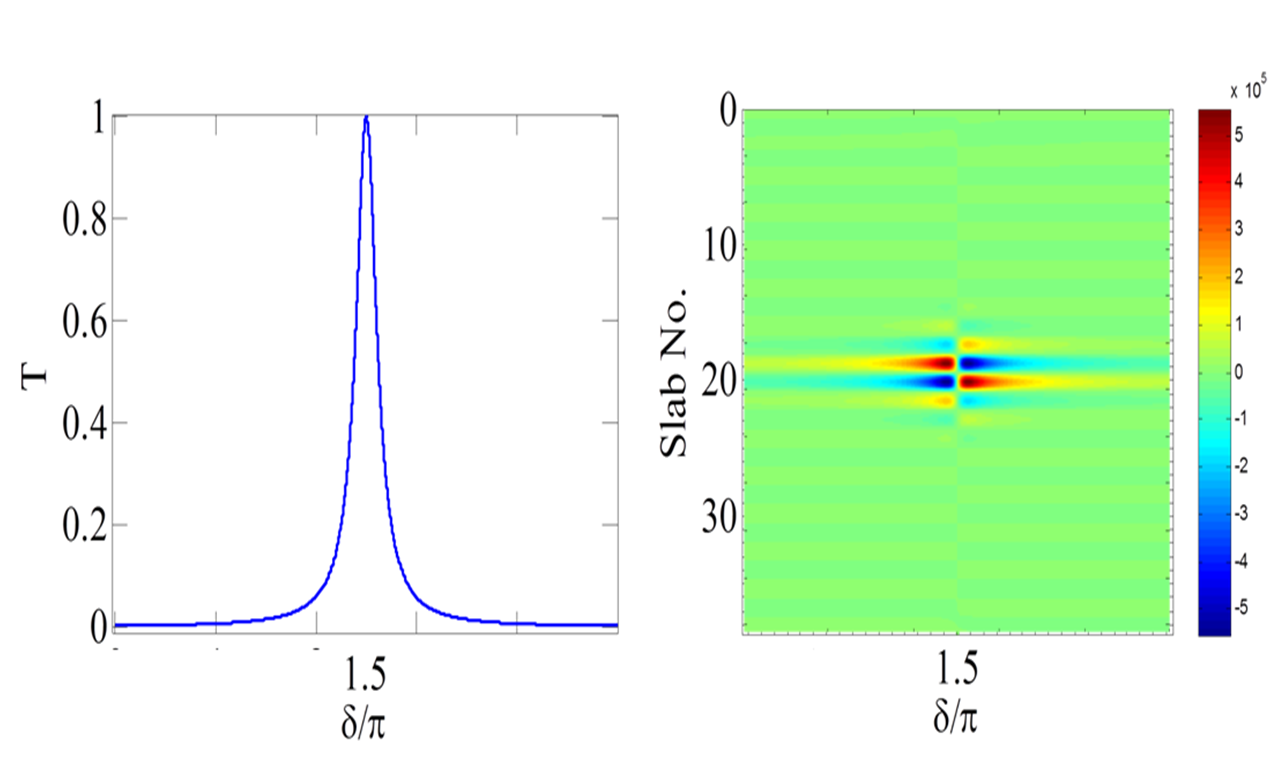}
\caption{A close-up on the transmission curve (left part) and the electric field intensity map (right part) for the resonant defect mode in the photonic crystal structure}
\label{figPCDMEmap}
\end{figure}
%\begin{figure}[ht]
%\includegraphics[width=11cm]{file.eps}
%\caption{Transmission curve, and electric amplitude map for the photonic crystal defect mode.}
%\end{figure}

A close up look at the defect mode transmission and its electric field amplitude map,
is depicted in figure \ref{figPCDMEmap}
The defect mode is indeed isolated spectrally and localized spatially.
However, if unintentional defects also exist, additional defect modes
can exist and may ``smear'' each other spatially and spectrally.

\section{Discussion}

 This chapter has presented the basic theoretical tools needed to describe
the spectral and transport properties of complex layered systems. These techniques are
quite standard in the fields of electrodynamics and quantum mechanics, but interesting
new effects emerge when applied to transport in complex layered structures.
The explicit numerical results for various types of layered samples show
that deterministic a-periodic structures can serve as a deterministic complex cavity,
hosting localized modes which are also narrow-band and spectrally isolated. In the Fibonacci dielectric structure, these modes also appear to be spatially self similar. This suggests promising applications for the physics of
spontaneous and stimulated emission, and these initial results deserve to be explored further, experimentally and numerically.

\medskip

This work was supported by US DOE under grant DE-FG02-92ER40716, and by the Israel Science Foundation Grant No.924/09. We thank E. Gurevich for helpful discussions and for relevant remarks.

\bibliographystyle{psp-rv-van}
\bibliography{psp-rv-sample}

\begin{thebibliography}{999}


\bibitem{pbg}
J. D. Joannopoulos, S. G. Johnson, J. N. Winn, and R. D. Meade,
{\it Photonic Crystals: Molding the Flow of Light}, (2nd Ed),
 (Princeton University Press, 2008); P. Lambropoulos, G. M. Nikolopoulos, T. R. Nielsen and S. Bay, ``Fundamental quantum optics in structured reservoirs'',  Rep. Prog. Phys. {\bf 63}, 455 (2000).

\bibitem{marcia}  E. Macia, ``The role of aperiodic order in science and technology'', Rep. Prog. Phys. {\bf 69}, 397 (2006);
E. L. Albuquerque, M. G. Cottam,
``Theory of elementary excitations in quasiperiodic structures'', Phys. Rep. {\bf 376}, 225 (2003);
 D.S. Wiersma, R. Sapienza, S. Mujumdar, M. Colocci, M. Ghulinyan, L. Pavesi, ``Optics of Nanostructured Dielectrics'', Journal of Optics A: Pure and Applied Optics, {\bf 7}, 190 (2005).

\bibitem{damanik} D. Damanik and A. Gorodetski, ``Spectral and Quantum Dynamical Properties
of theWeakly Coupled Fibonacci Hamiltonian'', Commun. Math. Phys. {\bf 305}, 221 (2011), and references therein.

\bibitem{erdos}
P. W. Anderson, ``Absence of diffusion in certain random lattices'', Phys. Rev.  {\bf 109}, 1492
(1958); P. Erd\"os and R. C. Herndon,
``Theories of electrons in one-dimensional disordered systems'',
Adv. Phys. {\bf 31}, 65 (1982);
P. Erd\"os, E. Liviotti and R. C. Herndon,
``Wave transmission through lattices, superlattices and layered media'',
J. Phys. D: Appl. Phys. {\bf 30}, 338 (1997);
C. D\'epollier, J. Kergomard and F. Lalo\"e, ``Anderson localisation of waves in acoustical random one-dimensional lattices'', Ann. Phys. (France) {\bf 11}, 457 (1986);
B. Kramer and A. MacKinnon, ``Localization: theory and experiment'', Rep. Prog. Phys. {\bf 56}, 1469
(1993);
M. V. Berry and S. Klein, ``Transparent mirrors: rays, waves and localization'', Eur. J. Phys. {\bf 18}, 222 (1997).

\bibitem{gredeskul}  I.M. Lifschits, S.A. Gredeskul and L.A. Pastur, {\it Introduction to the theory of
 disordered systems}, John Wiley (1988),

 \bibitem{luck} J.M. Luck, {\it Syst\`emes d\'esordonn\'es unidimensionnels}, Al\'ea Saclay (1992) and J.M. Luck, ``Cantor spectra and scaling of gap widths in deterministic aperiodic systems'', Phys. Rev. B {\bf 39},  5834 (1989).


\bibitem{ambook}
E. Akkermans and G. Montambaux,
{\it Mesoscopic Physics of Electrons and Photons},
(Cambridge University Press, 2007).


\bibitem{Frisch68} U. Frisch, ``Wave propagation in random media'', in {\it Probabilistic methods in
 applied mathematics}, A.T. Barucha-Reid eds., Vol.1, pp. 76-198, Academic Press (New York,
 1968); One may also consult:
 A. Ishimaru, {\it Wave propagation and scattering in random media}, 2 volumes, (Academic Press, 1978).


\bibitem{Budden}
K. G. Budden, {\it The Propagation of Radio Waves:
The Theory of Radio Waves of Low Power in the Ionosphere and Magnetosphere},
(Cambridge University Press, 1988).

\bibitem{nonlineaire} For preliminary works in the field, see: B. Dou\c cot and R. Rammal,
``On Anderson localization in nonlinear random media'', Europhys. Lett. {\bf 3}, 969 (1987);
P. Devillard and B. Souillard,
``Polynomially decaying transmission for the
nonlinear Schr\"odinger equation in a random medium'', J. Stat.
Phys. {\bf 43}, 423 (1986); For additional references, see: E. Akkermans, S. Ghosh and Z. Muslimani, ``Numerical study of one dimensional interacting Bose-Einstein condensates in a random potential'', J. Phys. B: At. Mol. Opt. Phys. {\bf 41},  045302, (2008).


\bibitem{tight} J. Callaway, {\it Quantum theory of solids} (chap. 6), Academic Press (1974).

\bibitem{3terms}  T. J. Rivlin,  {\it The Chebyshev Polynomials} (Wiley-Interscience, John Wiley and Sons, New York, 1974).
%\bibitem{master} ref. on master eq.

%\bibitem{rw} ref. on random walks


 \bibitem{avishai}
 Y. Avishai and Y. B. Band,
 ``One-dimensional density of states and the phase of the transmission amplitude'',
  Phys. Rev. B {\bf 32}, 2674 (1985).

\bibitem{konishi}
K. Konishi and  G. Paffuti,
{\it Quantum mechanics: a new introduction},
(Oxford University Press, 2009).

\bibitem{visser}
P. Boonserm and M. Visser,
``One dimensional scattering problems: a pedagogical presentation of the relationship between reflection and transmission amplitudes'',
Thai J. Math. Special Issue (2010), pages 83-97.

\bibitem{newton}
R. G. Newton,
{\it Scattering theory of waves and particles} (2nd ed.),
(Springer-Verlag, New York, 1982).

\bibitem{simon}
B. Simon,
{}``Notes on Infinite Determinants of Hilbert Space Operators'',
Adv. Math. {\bf 24}, 244 (1977).

\bibitem{kbs} M. G. Krein, Mat. Sb. (N.S.) 33, 597 (1953); Dokl.
Akad. Nauk SSSR 144, 268 (1962) [Sov. Math.-Dokl.
3, 707 (1962)]; M. Sh. Birman and M. G. Krein, Dokl.
Akad. Nauk SSSR 144, 475 (1962) [Sov. Math.-Dokl. 3,
740 (1962)]; I. C. Gohberg and M. G. Krein, {\it Introduction to the theory of linear nonselfadjoint operators}, (American Mathematical Society, 1969).

\bibitem{schwinger} J. Schwinger,
``On gauge invariance and vacuum polarization'', Phys. Rev. {\bf 82}, 664 (1951);
%``The Theory of Quantized Fields. V", Phys.\ Rev.\ {\bf 93}, 615 (1954);
%``The Theory of Quantized Fields. VI", Phys.\ Rev.\ {\bf 94}, 1362 (1954).
%``On gauge invariance and vacuum polarization'', Phys. Rev. {\bf 82} (1951) 664;
``The Theory of Quantized Fields. V'', Phys.\ Rev.\ {\bf 93}, 615 (1954);
``The Theory of Quantized Fields. VI'', Phys.\ Rev.\ {\bf 94}, 1362 (1954).

\bibitem{gold} J. S. Faulkner, ``Scattering theory and cluster calculations'', J. Phys. C: Solid State Phys. {\bf 10},  4661 (1977);  
M. L. Goldberger, ``On the second virial coefficient'', Phys. Fluids, {\bf 2}, 252 (1959).

\bibitem{thouless} D. J. Thouless, ``A relation between the density of states and range of
localization for one dimensional random systems'', J. Phys. C: Solid State Phys. {\bf 5}, 77 (1972); D. C. Herbert and R. Jones, ``Localized states in disordered systems'', J. Phys. C: Solid State  Phys. {\bf 4}, 1145 (1971).

\bibitem{fisher-lee}
D. S. Fisher and P. A. Lee,
``Relation between conductivity and transmission matrix'',
 Phys. Rev. B {\bf 23}, 6851 (1981).


\bibitem{gy} I.~M.~Gelfand and A.~M.~Yaglom,
{}``Integration In Functional Spaces And It Applications In Quantum Physics,''
J.\ Math.\ Phys.\ \textbf{1}, 48 (1960).

\bibitem{levit} S. Levit and U. Smilansky, {}
``A theorem on infinite products of eigenvalues of Sturm-Liouville type operators'',
Proc. Am. Math. Soc. \textbf{65}, 299 (1977).

\bibitem{kleinert} H.~Kleinert,
{\it Path Integrals in Quantum Mechanics, Statistics, Polymer Physics, and Financial Markets},
(World Scientific, Singapore, 2004).

\bibitem{dets} G.~V.~Dunne,
``Functional determinants in Quantum Field Theory'',
J.\ Phys.\ A \textbf{41}, 304006 (2008), {[}arXiv:hep-th/0711.1178].

\bibitem{berry-mount}
M. V. Berry and K. E.  Mount,
``Semiclassical approximations in wave mechanics'', Reps.\  Prog.\ Phys.\ {\bf 35}, 315 (1972).

\bibitem{schiff}
J. Schiff and S. Shnider,
``A natural approach to the numerical integration of Riccati differential equations'',
SIAM J. Numer. Anal. {\bf 36}, 1392 (1999);
C-C. Chou and R. E. Wyatt,
``Computational method for the quantum Hamilton-Jacobi equation: One-dimensional scattering problems'',
Phys. Rev. E {\bf 74}, 066702 (2006).




\bibitem{Born-wolf}
M. Born and  E. Wolf,
{\it Principles of Optics: Electromagnetic Theory of Propagation, Interference and Diffraction of Light},
(Cambridge University Press, 1999).

%\bibitem{Kohmotocanonic} Kohomoto

%\bibitem{banno}
%I. Banno, D. Kaneko and K. Fujima,
%``Determinant of Wave Operator as a Measure for Resonance
%and Bound States in One-Dimensional Potential Scattering System'',
%Optical Rev. {\bf 13}, 249 (2006).

\bibitem{quasi}
M. Kohmoto, L. P. Kadanoff and C. Tang, ``Localization Problem in One Dimension: Mapping and Escape'',
Phys. Rev. Lett. {\bf 50}, 1870 (1983);
S. Ostlund, R. Pandit, D. Rand, H. J. Schellnhuber and E. Siggia,
``One-Dimensional Schršdinger Equation with an Almost Periodic Potential'',
Phys. Rev. Lett. {\bf 50}, 1873 (1983).

\bibitem{kohomotoothers}
 W. Gellermann, M. Kohmoto, B. Sutherland and P. C. Taylor,
 ``Localization of light waves in Fibonacci dielectric multilayers'', Phys. Rev. Lett. {\bf 72}, 633 (1994);
 M. Kohmoto, B. Sutherland  and C. Tang,
 ``Critical wave functions and a Cantor-set spectrum of a one-dimensional quasicrystal model'',  Phys. Rev. B {\bf 35}, 1020 (1987);  
 M. Kohmoto, B. Sutherland  and K. Iguchi,
 ``Localization of optics: Quasiperiodic media'',  Phys. Rev. Lett. {\bf 58}, 2436 (1987).


\bibitem{esaki}
R. Tsu and L. Esaki,
``Tunneling in a finite superlattice'',
Appl. Phys. Lett. {\bf 22}, 562 (1973).
%; http://dx.doi.org/10.1063/1.1654509


\bibitem{damanik2} D. Damanik, M. Embree, A.  Gorodetski, S. Tcheremchantsev,
``The fractal dimension of the
spectrum of the fibonacci hamiltonian'',  Commun. Math. Phys. {\bf 280}, 499 (2008);
J. Bellissard, A.  Bovier, J.-M. Ghez,
``Gap labelling theorems for one-dimensional discrete
schršdinger operators'',  Rev. Math. Phys. {\bf 4}, 1 (1992).

\bibitem{dalnegro1} L. Dal Negro, C. J. Oton, Z. Gaburro, L. Pavesi, P. Johnson, A. Lagendijk, R. Righini, M. Colocci and D. S. Wiersma,
``Light Transport through the Band-Edge States of Fibonacci Quasicrystals'',  Phys. Rev. Lett. {\bf 90},  055501 (2003);
L. Dal Negro, M. Stolfi, Y. Yi, J. Michel, X. Duan, L. C. Kimerling, J. LeBlanc and J. Haavisto,
``Photon band gap properties and omnidirectional reflectance in Si/SiO2 ThueÐMorse quasicrystals'',  Appl. Phys. Lett. {\bf 84}, 5186 (2004).

\bibitem{dalnegro2} A. Gopinath, S. V. Boriskina, B. M. Reinhard and L. Dal Negro, ``Deterministic aperiodic arrays of metal nanoparticles for surface-enhanced Raman scattering (SERS)'', Optics Express, {\it 17}, 3741 (2009);
 L. Dal Negro, N. Feng and A. Gopinath,
 ``Electromagnetic coupling and plasmon localization in deterministic aperiodic arrays'',
 J. Opt. A: Pure Appl. Opt. {\bf 10}, 064013  (2008).

\bibitem{tmatrixrho}
J.M. Bendickson, J.P. Dowling, and M. Scalora,
``Analytic expressions for the electromagnetic mode density in finite, one-dimensional, photonic band-gap structures'', Phys. Rev. E, {\bf 53}, 4107 (1996).




\end{thebibliography}
%
%%\printindex[aindx]                 % to print author index
%\printindex                         % to print subject index
\end{document}